# Searching for Spectroscopic Signatures of Ongoing Quenching in SDSS Galaxies

Andrea Weibel[1,2], Enci Wang[1], and Simon J. Lilly[1]
[1] Department of Physics, ETH Zurich, Wolfgang-Pauli-Strasse 27, CH-8093 Zurich, Switzerland; andrea.weibel@unige.ch, enci.wang@phys.ethz.ch
[2] Departement d'Astronomie, Université de Genève, 51 Chemin Pegasi, CH-1290 Versoix, Switzerland

## Abstract

In this paper, we estimate the "star formation change parameter," $SFR_{79}$, which characterizes the current SFR relative to the average during the last 800 Myr, for ∼300,000 galaxies selected from the Sloan Digital Sky Survey (SDSS). The goals are to examine, in a much larger and independent sample, the trends previously reported in a sample of star-forming (SF) MaNGA galaxies and also to search for spectroscopic signatures of ongoing quenching in the so-called "Green Valley," which is generally believed to contain galaxies that are migrating from the SF population to the quenched population of galaxies. Measuring $SFR_{79}$ for our large sample of SDSS galaxies, we first confirm the basic results of SF galaxies published by Wang & Lilly. We then discuss in detail the calibration and meaning of $SFR_{79}$ for galaxies that are well below the SFMS and establish the expected statistical signature of systematic ongoing quenching from modeling the $z \sim 0$ quenching rate of the SF population. We conclude that it is not possible at present to establish unambiguous observational evidence for systematic ongoing quenching processes, due to limitations both in the noise of the observational data, in particular in the measurements of $H\delta$ absorption, and in the calibration of $SFR_{79}$, as well as biases introduced by the necessity of selecting objects with significant $H\alpha$ emission. We do however see plausible indications of ongoing quenching, which are quantitatively consistent with expectations from "growth+quenching" models of galaxy evolution and a typical e-folding timescale for quenching of ∼500 Myr.

*Unified Astronomy Thesaurus concepts:* Galaxy quenching (2040); Galaxy evolution (594); Star formation (1569); Green valley galaxies (683); Galaxies (573); Galaxy spectroscopy (2171)

## 1. Introduction

Galaxies over a broad range of cosmic epochs can broadly be divided into two distinct populations based on their distribution in the star formation rate–stellar-mass ($SFR$–$M_*$) plane: active, star-forming (SF) galaxies and passive, quenched galaxies (e.g., Strateva et al. 2001; Baldry et al. 2004; Bell et al. 2004; Blanton et al. 2005; Faber et al. 2007; Wetzel et al. 2012; Wang et al. 2018a). This bimodality is seen out to a redshift of at least 2.5 (e.g., Bundy et al. 2006; Martin et al. 2007; Muzzin et al. 2012). SF galaxies form a relatively tight and slightly sublinear sequence (e.g., Brinchmann et al. 2004; Daddi et al. 2007; Noeske et al. 2007; Pannella et al. 2009; Elbaz et al. 2011; Stark et al. 2013; Renzini & Peng 2015; Boogaard et al. 2018; Wang et al. 2019), which is often called the SF main sequence (SFMS). In contrast, the passive population is characterized by little or no star formation and by structural morphologies that are more dominated by spheroids (e.g., Baldry et al. 2004; Li et al. 2006; Wuyts et al. 2011; Muzzin et al. 2013; Barro et al. 2017). The mass functions of these two populations are noticeably different, with passive galaxies dominating the galaxy mass function at high masses (e.g., Peng et al. 2010).

This galaxy bimodality is generally interpreted in terms of a scenario in which, after formation, galaxies reside on the SFMS, forming stars and continually increasing their stellar mass as gas is accreted from the growing halo. At some point, however, a given galaxy "quenches," i.e., its SFR drops by a factor of 10 or more, and it joins the passive population. This quenching process may be due to a number of different physical processes operating in and around both satellite galaxies and the central galaxies of dark matter halos. Peng et al. (2010; see also Peng et al. 2012) introduced a useful distinction between different quenching channels of "mass quenching," which limits the mass of galaxies as they approach the characteristic Schechter $M^*$, and "environment-" or "satellite-quenching" which operates only on satellite galaxies and is more or less independent of their stellar mass.

The region between the dominant blue SF and the red quenched populations is often called the Green Valley (GV; e.g., Salim et al. 2007; Wyder et al. 2007; Schawinski et al. 2014; Smethurst et al. 2015; Mahoro et al. 2017; Belfiore et al. 2018; Nogueira-Cavalcante et al. 2018; Wang et al. 2018a). The original definition of the GV in terms of a relative paucity of galaxies with intermediate photometric colors is easily translatable to an equivalent feature in the distribution of SFRs at a given mass, where there is seen to be a paucity of galaxies with intermediate values of specific SFR (sSFR; defined as $SFR/M_*$) in logarithmic space.

It is therefore often assumed that galaxies in the GV are transitioning from the SFMS to the quenched population, i.e., that they are objects that are currently undergoing a quenching process. However, the time development of the individual SFRs of GV galaxies is hard to determine from observations, as it requires measurements of both the current SFR and the SFR in the recent past. Wang & Lilly (2020a), hereafter WL20, recently developed a parameter that characterizes the change of the star formation over gigayear timescales for SF galaxies. In this paper, we will explore the applicability of this approach to galaxies below the SFMS and search for direct spectral evidence of ongoing quenching processes.

In order to characterize quenching as a process, it is convenient to assume that the SFR in a quenching galaxy is declining exponentially with time. This then yields the e-folding timescale of the declining SFR, $\tau_Q$, as a useful parameterization of the process (e.g., Wetzel et al. 2013; Hahn et al. 2017). Several authors have tried to statistically estimate or constrain $\tau_Q$, but the results have not yielded consistent or convergent results (Wetzel et al. 2013; Schawinski et al. 2014; Yesuf et al. 2014; Peng et al. 2015; Hahn et al. 2017; Smethurst et al. 2018; Trussler et al. 2020). For instance, by constructing a model that statistically tracks SFHs and quenching of central galaxies, Hahn et al. (2017) find that the median quenching timescale decreases as a function of $M_*$ from $\tau_Q = 1.2$ Gyr at $M_* = 5 \times 10^{10} M_\odot$ to $\tau_Q = 400$ Myr at $M_* = 2 \times 10^{11} M_\odot$. A similar trend is found specifically for satellites by Wetzel et al. (2013) but with a much shorter quenching timescale, where the typical $\tau_Q$ is claimed to decrease from $\tau_Q \approx 800$ Myr at $M_* = 6 \times 10^9 M_\odot$ to $\tau_Q \approx 200$ Myr at $M_* = 10^{11} M_\odot$. On the contrary, with an analysis of the stellar metallicity in local galaxies, Peng et al. (2015) claimed that strangulation is the primary mechanism responsible for quenching and found a typical transitioning timescale from the SF to the quenched population of as long as 4 Gyr. Their approach was further developed and improved in Trussler et al. (2020), who convert their quenching durations to e-folding timescales. They find typical timescales of $\tau_Q = 2$–4 Gyr in their closed-box models, which assume a pure cessation of the gas supply, followed by a sudden removal of the remaining gas. If they assume a leaky-box model that incorporates continuous outflows, the typical timescale decreases to $\tau_Q \sim 1$ Gyr. They further claim that local GV galaxies show slightly longer quenching timescales compared to passive galaxies with typical values $\tau_Q = 1.5$–2.5 Gyr, indicating that quenching is slower in the local universe than at higher redshifts. It is clear that further observational work is required to clarify the timescales of the suppression of star formation as well as the underlying physical mechanisms.

To directly see whether or not GV galaxies are quenching their star formation in real time and to determine the timescale of this suppression, it is necessary to constrain the time-evolution of their SFRs on extended timescales. It is not of course possible to monitor any individual galaxy over any interesting timescale—they are all seen at single snapshots of their evolution. It is however possible to characterize the SFR of a given galaxy *averaged* over different timescales. WL20 developed a framework for that based on the optical spectral features of H$\alpha$ emission, H$\delta$ absorption, and the 4000 Å break. In principle (although see the discussion in that paper), the measurement of these three spectral features should be independent of the effects of reddening by dust.

The H$\alpha$ emission line from H II regions traces the recent SFR within the last 5 Myr (SFR$_{5\text{Myr}}$), while the H$\delta$ absorption feature roughly traces the SFR averaged over the last 800 Myr, SFR$_{800\text{Myr}}$ (e.g., Balogh et al. 1999; Kauffmann et al. 2003; Li et al. 2015; Wang et al. 2018a). A "star formation change parameter" can then be defined as the ratio of the SFRs averaged on these two timescales, i.e., SFR$_{5\text{Myr}}$/SFR$_{800\text{Myr}}$.

Using the SFR$_{79}$, WL20 investigated the variability in the SFR of SF galaxies. They found, among other things, that galaxies with a recent temporal enhancement (or suppression) in their overall SFR have enhanced (or suppressed) star formation at all galactic radii. In addition, galaxies, or regions of galaxies, with short gas depletion times, i.e., with high star formation efficiency, appear to undergo larger-amplitude temporal variations in their SFRs. Exploring this further, Wang & Lilly (2020b) constrained the temporal power spectrum of the sSFR of SF galaxies. The results of both WL20 and Wang & Lilly (2020b) are consistent with the dynamical response of a gas-regulator system (Lilly et al. 2013) to a time-varying inflow, as previously proposed in Wang et al. (2019).

In the present work, our first goal is to test those results using a much larger data set obtained from the Sloan Digital Sky Survey (SDSS). We then apply the star formation change parameter concept to galaxies that lie significantly below the SFMS and discuss in detail the limitations and caveats of this procedure. In order to search for direct evidence of systematic ongoing quenching, we carefully investigate the expected impact of a population of quenching galaxies on the observed distribution of the star formation change parameter. Finally, we try to address the question of whether the signature of a population of quenching galaxies can be identified, and if so, on what timescale(s) these galaxies are decreasing their (s)SFR.

Our approach of directly converting observed spectral indices to SFRs on different timescales and using this to constrain the time-variability or evolution of the SFR is somewhat similar to the approach in Martin et al. (2017), which is further applied in the framework of quenching in Darvish et al. (2018). In their work, photometric broadband colors between the *FUV*, *NUV*, *u*, *g*, *r*, *i*, and *z* bands are used in addition to the H$\delta$ absorption Lick index and $D_n4000$ to constrain various physical parameters, including the so-called star formation acceleration SFA, which is defined as the time derivative of the $NUV - i$ color, i.e., the change in this color over the past 300 Myr. They find evidence for ongoing net quenching in low-mass, low-sSFR galaxies as well as enhanced quenching in high-mass centrals compared to satellites and in galaxies that host an active galactic nucleus (AGN). Further, they find indications for short typical quenching and bursting timescales of $\sim 300$ Myr assuming exponentially declining SFHs, and Darvish et al. (2018) argue that quenching was generally stronger at $z \sim 1$ than it is in the local universe.

The layout of the paper is as follows. In Section 2, we present the data that are used in the present work, including the sample selection, an improved method to estimate the H$\delta$ absorption index, and a refined calibration of the star formation change parameter SFR$_{79}$ as compared to WL20. We examine a number of issues associated with estimating SFR$_{79}$ for galaxies below the SFMS. In Section 3 we define a sample of SFMS galaxies and establish broad consistency with the results published in WL20 and Wang & Lilly (2020b). In Section 4, we examine the SFR$_{79}$ for galaxies lying significantly below the SFMS. We derive the expected effect of systematic ongoing quenching on the distribution of SFR$_{79}$ and search for direct signatures of ongoing quenching. Finally, we provide indicative estimates of what the typical quenching timescales could be. We summarize our main conclusions in Section 5.

Throughout this paper, we use the following shorthand notation: The SFR averaged over the last 5 Myr, SFR$_{5\text{Myr}}$, is denoted as SFR$_7$ (because 5 Myr $\approx 10^7$ yr); that averaged over the last 800 Myr, SFR$_{800\text{Myr}}$, as SFR$_9$ (because 800 Myr $\approx 10^9$ yr); and the ratio of the two, SFR$_{5\text{Myr}}$/SFR$_{800\text{Myr}}$, as SFR$_{79}$, consistent with WL20. We continue using their notation for consistency but stress here that SFR$_{79}$ is a ratio of two SFRs and therefore unitless and not itself an SFR. When computing distance-dependent quantities, we assume a flat cold dark matter cosmology with $\Omega_m = 0.27$, $\Omega_\Lambda = 0.73$, and $H_0 = 70$ km s$^{-1}$ Mpc$^{-1}$.

## 2. Data

### 2.1. Selection of the SDSS Sample

We select our main galaxy sample from the MPA-JHU catalog[3] (Kauffmann et al. 2003; Brinchmann et al. 2004; Tremonti et al. 2004) of the seventh data release of the Sloan Digital Sky Survey (SDSS DR7; Abazajian et al. 2009) where 927,552 extragalactic objects are listed. In the SDSS, spectra are obtained through 3″ diameter fibers, which are centered on galactic centers. The wavelength coverage is 3800–9200 Å.

We adopt the stellar masses (both integrated and just within the fiber aperture) from the MPA-JHU catalog in which they were determined based on fits to the photometry. These mass estimates agree well with the stellar masses from Kauffmann et al. (2003) based on spectral indices. It should be noted that these mass estimates represent the mass of living or shining stars. However, in our calibrator, which we will introduce in Section 2.3.1, the stellar mass is defined as the integral of the SFH. Therefore, for consistency, we correct the MPA-JHU mass to represent the integrated SFH by taking the fraction of the stellar mass that is returned to the interstellar medium through winds and supernova explosions into account. We adopt the return fraction to be 0.4 for all galaxies for the Chabrier (2003) initial mass function (IMF; Vincenzo et al. 2016).

Our basic preselection criteria for this study are as follows: (1) $z < 0.2$, (2) the object is not a duplicate, and (3) it can be matched with the UPenn PhotDec catalog (Meert et al. 2015). The third criterion implies in turn that we adopt all the selection criteria of the UPenn PhotDec catalog. These are specified in Meert et al. (2015) to be (a) the extinction-corrected Petrosian magnitude in the r-band is between 14 and 17.77 mag, and (b) the object is identified to be a galaxy based on the spectroscopy and the photometry. The limit at the bright end is set to exclude very nearby and large objects, which are either too well resolved to be fitted with a standard light profile or are split into multiple objects in the SDSS catalog. The limit at the faint end is adopted from Strauss et al. (2002) to be the limit on the completeness of the sample. An additional ≈5000 objects are removed because they have $z < 0.005$, very low surface brightness, or data quality issues (see details in Meert et al. 2015).

Overall, this yields a sample consisting of 619,960 galaxies. Of these, some 4830 galaxies (less than 0.8%) cannot be used for the subsequent analysis, because either the corresponding spectra are not available from SDSS DR12 (333 cases), or the data file is truncated (21 cases), or, most often, the fitting routine applied (see Section 2.2) fails to provide a reasonable fit to the continuum of the spectrum (4476 cases). This leaves us with a final sample of 615,130 galaxies.

We cross-match this sample with the SDSS DR7 group catalog published by Yang et al. (2007) in order to distinguish between centrals and satellites. The group catalog consists of 639,359 objects, 555,594 of which are matched with the sample investigated here. We adopt the most massive galaxy in each group to be the central galaxy, as recommended. Yang et al. (2007) tested their group finder on a mock galaxy redshift survey that mimics the SDSS DR4 and specify the fraction of galaxy groups in the mock survey whose central galaxy has actually been identified as the central galaxy of its group to be around 90% for $M_{\rm halo} > 10^{12} M_\odot$. The vast majority of the remaining 10% of central galaxies are wrongly classified to be satellites. We note that this effect alone would cause a ~28% contamination of the satellite population in our final sample as it contains 2.8 times more centrals than satellites. In reality, the contamination is likely to be even larger because the 90% quoted above does not capture the full uncertainty of the group finder (see, for example, Knobel et al. 2009 for a more detailed discussion of impurity and contamination effects in group-finding algorithms).

On the left of Figure 1, we show the mass–redshift distribution of the full sample. The color coding represents the number density $\rho_N$, defined as the number of objects in a given grid cell divided by the total number of objects and the area of the cell in parameter space.

Since the sample is flux limited, it is biased toward objects with a low mass-to-light ratio, i.e., toward bright, SF objects (e.g., Faber & Gallagher 1979; Girardi et al. 2000; Bell et al. 2003; Cappellari et al. 2006). Below, we will wish to quantitatively analyze the relative number of SF and currently quenching galaxies. Therefore, we assign a mass-dependent redshift cut to the sample galaxies to ensure that, at each mass, there is no bias against high mass-to-light objects and further that the sample is statistically complete out to some limiting (mass-dependent) redshift.

To do this, we start by focusing on objects with a low SFR and thus typically high mass-to-light ratio. Specifically, we select galaxies that are between 0.9 and 1.1 dex below the SFR$_7$-based SFMS (see definition in Section 3.1). These objects are just below the lower boundary in EW(H$\alpha$) (and thus sSFR$_7$), above which we argue the calibration of SFR$_{79}$ to be meaningful (see Section 2.3.3). We then split these galaxies into 20 mass bins with a width of 0.1 dex. For each mass bin, we define a cut in redshift, below which the sample galaxies are complete for r-band magnitudes less than 17.77 mag. To avoid confusion, we note that the subsample considered here is only used to define reasonable mass-dependent redshift cuts and not to do any scientific analysis. The resulting mass-dependent redshift cuts are indicated by the stepped black lines in Figure 1.

To better illustrate and justify these cuts, we show the $M_*$–$z$ distribution of the mentioned subsample, overlapped with the redshift cuts on the right of Figure 1. Those galaxies are expected to have a higher mass-to-light ratio on average than the SF galaxies, which shifts the lower boundary of the $M_*$–$z$ distribution up and to the left. We argue that the redshift cuts are conservative enough to avoid any bias for objects with SFRs representative of those found as far as 1 dex below the SFMS.

It should be noted that galaxies with high stellar masses are inevitably "overrepresented" in our sample because of the greater accessible volumes for these. This is not a concern in the present work, since we will usually be interested in the properties of galaxies as a function of stellar mass and not in the relative numbers of high- and low-mass galaxies. We will often work in four mass bins with a width of 0.5 dex in $\log(M_*/M_\odot)$, respectively. Those four bins are indicated by the solid horizontal lines in Figure 1.

After the application of the mass–redshift cuts, we are left with 274,794 galaxies. Of these, 192,195 (69.9%) are identified as centrals, 68,528 (24.9%) are identified as satellites, and the remaining 14,071 (5.1%) cannot be identified as either centrals

---

[3] https://wwwmpa.mpa-garching.mpg.de/SDSS/DR7

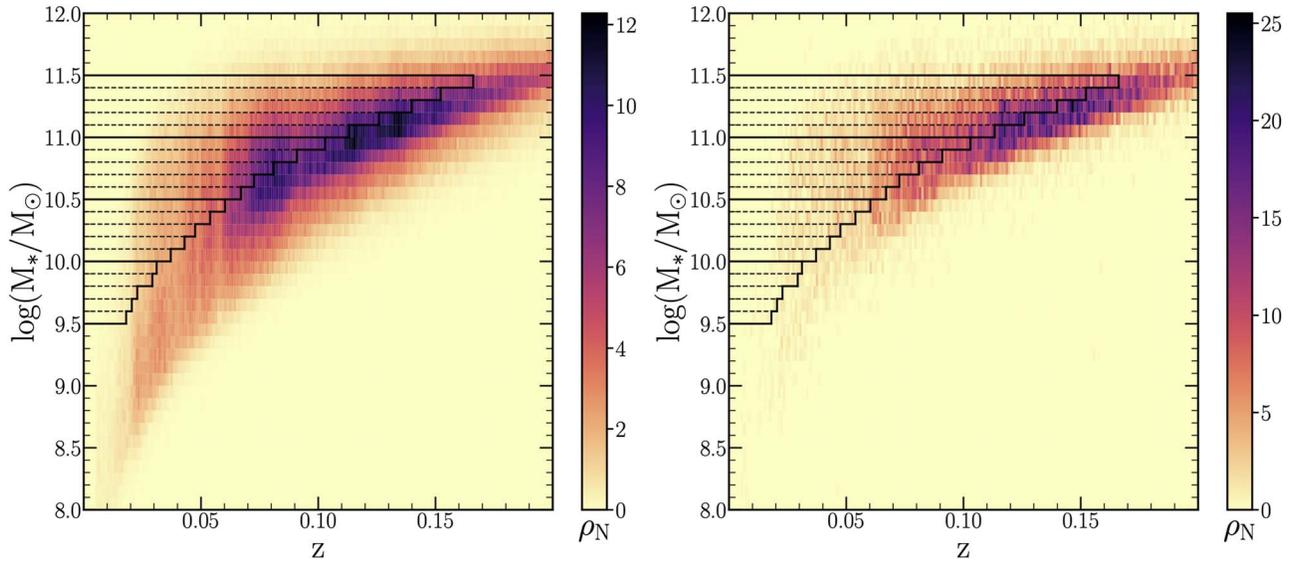

**Figure 1.** Left panel: mass–redshift distribution of the full sample consisting of 615,130 galaxies. Right panel: mass–redshift distribution for a subsample located within 0.9–1.1 dex below the SFR$_7$-based SFMS (see Section 3.1). The $M_*$–$z$ plane is split into grid cells with a width of 0.1 dex in mass and 0.001 in redshift. Each grid cell is colored according to the number density $\rho_N$ of objects in it. $\rho_N$ is defined as the number of objects in a given grid cell divided by the total number of objects and the area of the cell in parameter space. The horizontal solid lines indicate four mass bins with a width of 0.5 dex, which we will use throughout the paper, the dashed lines indicate mass bins with a width of 0.1 dex in each of which we defined a cut in redshift as explained in the text to avoid being biased toward objects with a low mass-to-light ratio. Since this is of particular concern for quenching or quenched objects (i.e., objects with a low sSFR), the plot on the right shows the exact same mass and redshift bins but it only shows objects that are between 0.9 and 1.1 dex below the SFMS in sSFR$_7$, which is just below the lowest sSFR's that we will investigate in this paper. Those objects were used to define the redshift cuts in the first place.

or satellites because they are not listed in the group catalog of Yang et al. (2007).

### 2.2. Measurements of the Spectral Features

The star formation change parameter SFR$_{79}$ is based on three observational quantities: the equivalent width of the H$\alpha$ emission line (EW(H$\alpha$)), the Lick index of H$\delta$ absorption (EW(H$\delta_A$)), and the size of the 4000 Å break ($D_n$4000). Conveniently, EW(H$\alpha$) is sensitive to the SFR over the last ∼5–10 Myr, EW(H$\delta_A$) to the SFR over the last ∼1 Gyr, and $D_n$4000 traces the SFR over an even longer timescale of ∼1–10 Gyr, albeit with a substantial metallicity dependence (see WL20). The bandpass for the computation of $D_n$4000 is defined in Balogh et al. (1999) as [3850, 3950] and [4000, 4100] Å. The three bandpasses used for the Lick index EW(H$\delta_A$) are [4083.50, 4122.25], [4041.60, 4079.75], and [4128.50, 4161.00] Å (Worthey & Ottaviani 1997). These are consistent with the definitions used in the MPA-JHU catalog. Since the three parameters are each measured over a very short wavelength interval, they are in principle insensitive to the effects of dust attenuation. However, as discussed in WL20, dust may still have an effect if different components of the galaxy in question (i.e., the stellar continuum and the nebular emission) have different levels of dust extinction (e.g., Calzetti et al. 2000; Moustakas & Kennicutt 2006; Wild et al. 2011; Hemmati et al. 2015). After measuring the three spectral features, we therefore perform an empirical correction adopted from WL20 to account for differential dust extinction between the emission lines and the continuum (see details in Section 2.6 of WL20).

One of the main difficulties in this work is to accurately measure the EW(H$\delta_A$) for spectra with a relatively low signal-to-noise ratio (S/N). This is because the H$\delta$ absorption line is usually contaminated by H$\delta$ emission that must be accurately subtracted. In principle, it would be possible to fit both absorption and emission simultaneously, provided that they did not have the same line profile. We performed an initial spectral fitting with the penalized PiXel-Fitting (pPXF) code (Cappellari & Emsellem 2004; Cappellari 2017) using 150 stellar population template spectra from Falcón-Barroso et al. (2011), which are based on the MILES library (Sánchez-Blázquez et al. 2006). We correct for foreground Galactic extinction by adopting the reddening map from Schlegel et al. (1998) and the dust extinction curve from O'Donnell (1994). In pPXF, the spectral contribution of the emission lines can in principle be obtained by subtracting the stellar contribution from the observed spectra and simultaneously fitting all the emission lines subject to an accurate knowledge (and modeling) of the instrumental line spread function.

Indeed, the EW(H$\alpha$) parameter is easily obtained directly from this fitting process. We note here however that 43,876 galaxies (16%) show an unreasonable velocity dispersion in their H$\alpha$ emission lines (i.e., the dispersion of the fitted Gaussian is outside the range $\sigma \in (0.5, 7.5)$ Å, which roughly corresponds to $\sigma \in (50, 800)$ km s$^{-1}$) or extremely low H$\alpha$ fluxes (i.e., lower than their nominal uncertainties). Those galaxies, likely passive galaxies without significant H$\alpha$ emission, are ignored completely in the following analysis unless otherwise stated.

In principle, we can use the same fitted spectra to subtract the emission lines in the respective wavelength intervals from the data and measure both $D_n$4000 and EW(H$\delta_A$) without the contribution of emission lines. However, it became clear that the measurements of EW(H$\delta_A$) obtained in this way were unsatisfactory, at least for the generally low S/N of the SDSS spectra. Unfortunately, accurate measurements of EW(H$\delta_A$) are critical when deriving SFR$_{79}$ because its error almost always dominates the observational uncertainty of SFR$_{79}$.

To gain insight into the performance of the fitting procedure at low S/N, we first construct a set of representative spectra with extremely high S/N by stacking many individual SDSS spectra that were matched in EW(H$\alpha$), stellar mass, and the S/N in the H$\delta$ continuum. We then randomly select 100 of these stacked spectra and measure their EW(H$\delta_A$) with the spectral fitting method. We adopt these measured values to be the "true" values of the representative spectra. For each of these 100 stacked spectra, we then produce 25 realizations at each of 15 different S/N levels by adding uncorrelated noise to the original high-S/N stacked spectrum. We are therefore left with 37,500 noisy mock spectra, spanning a wide range of S/N. We then remeasure the EW(H$\delta_A$) using the same spectral fitting method and compare the values to the corresponding "true" values that have been obtained from the undegraded spectra.

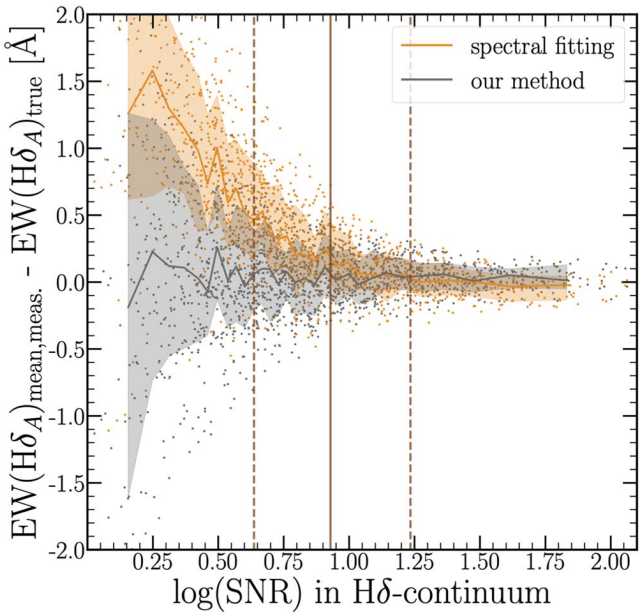

**Figure 2.** Examination of the two approaches in measuring the EW(H$\delta_A$) for spectra of different S/N. The dots represent the deviation of the mean EW(H$\delta_A$) of 25 randomizations of a set of 1500 stacked spectra with different noise levels added in from the true value respectively for the spectral fitting method in orange and for our new method to derive the EW(H$\delta_A$) in gray, as a function of the log(S/N) in the H$\delta$ continuum of the perturbed stacks. The orange and gray lines represent the median deviation of 50 consecutive mean values, and the shaded regions represent the region from the 16th to the 84th percentile, respectively. The vertical solid brown line indicates the median S/N, and the dashed brown lines represent the 5th and 95th percentile of the S/N in our sample.

As shown by the orange curve in Figure 2 there is a clear and rather large systematic deviation in the mean measured EW(H$\delta_A$) at low S/N. The spectral fitting method significantly overestimates the EW(H$\delta_A$) at low S/N[4] (S/N < 10), and this systematic deviation becomes more significant with decreasing S/N. For reference, the median S/N of our sample at the wavelength of H$\delta$ is 8.5 with 90% of the spectra between 4.3 and 17.2 (see the brown lines in Figure 2).

An alternative approach is to estimate the strength of the H$\delta$ emission from the observed emission-line fluxes of H$\alpha$ and H$\beta$ since these lines are stronger and generally observed at a much higher S/N. The intrinsic flux ratios between the Balmer lines should be fixed, (H$\alpha$/H$\beta$)$_{int}$ = 2.86 and (H$\delta$/H$\beta$)$_{int}$ = 0.259, under the assumption of case B recombination, a temperature of $T = 10^4$ K, and an electron density of $n_e = 100$ cm$^{-3}$ (Osterbrock 1989). Adopting the extinction curve of O'Donnell (1994), one can then establish the extinction based on the Balmer decrement (e.g., Domínguez et al. 2013) and thereby predict the flux of the H$\delta$ emission line. We can therefore subtract this predicted contribution from the EW(H$\delta_A$) Lick index that is obtained from the raw observed spectra. We note that no dust attenuation correction is performed for galaxies with (H$\alpha$/H$\beta$)$_{obs}$ < 2.86. For galaxies with an S/N < 3 in either the H$\alpha$ or the H$\beta$ emission line, we apply a dust correction based on a median $E(B-V)$ from galaxies with low H$\alpha$ (and H$\beta$) emission but S/N > 3 in both lines.

---

[4] For the same reason, we do not simply adopt the measurements of EW(H$\delta_A$) from the MPA-JHU catalog.

We tested this new approach in the same way as above. Specifically, we applied the new method to the same 37,500 mock spectra and then compared the derived measurements to the "true" values previously obtained from the spectral fitting method applied to the high-S/N spectra.

As can be seen through the gray curve in Figure 2, this new method is both completely consistent with the standard fitting procedure at high S/N and largely eliminates the systematic bias at low S/N. It improves the EW(H$\delta_A$) measurements and appears to be quite stable for spectra of low S/N. Therefore, in the following, all measurements of EW(H$\delta_A$) are based on this new method.

In addition, the analysis outlined above provides a way to quantitatively estimate the observational uncertainties of the final EW(H$\delta_A$) as a function of the S/N. This uncertainty is taken to be the scatter among the EW(H$\delta_A$) measurements of the mock spectra at a given S/N, i.e., the width of the gray-shaded region in Figure 2. We therefore assign this empirical uncertainty in EW(H$\delta_A$) to each individual galaxy based on its nominal S/N in the H$\delta$ continuum.

### 2.3. Estimation of the SFR Parameters

We can obtain the SFR$_7$ straightforwardly from the H$\alpha$ luminosity by adopting the relation from Kennicutt & Robert (1998) and using the Chabrier (2003) IMF (for consistency, see below). We obtain the star formation change parameter, SFR$_{79}$ using a similar but improved method to that in WL20, as detailed below. Combining this with the estimate of SFR$_7$ we can then also obtain the SFR$_9$ for each individual galaxy.

#### 2.3.1. Estimation of SFR$_{79}$

Here, we will briefly summarize the method, referring the readers to WL20 for details, but highlight our further development of it.

We first construct millions of mock SFHs of galaxies that are designed to cover as comprehensively as possible the range of possible SFHs of galaxies in the universe. As described in WL20, these mock SFHs consist of a smooth underlying SFH on which short-term stochastic variations are superposed. We then generate synthetic spectra of these mock galaxies at the present epoch based on stellar population models of the corresponding stellar metallicity (assuming the mass–metallicity relation from Zahid et al. 2017), using the Flexible Stellar Population Synthesis code (FSPS; Conroy et al. 2009). In this process, we adopt the MILES stellar library (Sánchez-Blázquez et al. 2006; Falcón-Barroso et al. 2011), a Chabrier (2003) IMF, and the Padova isochrones (e.g., Bertelli et al. 1994, 2008). Using simple stellar populations as the ionizing source for the gas clouds, CLOUDY (Ferland et al. 2013) predicts the thermal, ionization, and chemical structure of the clouds and further provides the resulting spectrum of the diffuse emission by simulating physical conditions within a gas cloud. In the FSPS model, the ionizing radiation is assumed to be produced by a point source at the center of a cloud with a spherical shell geometry and a constant gas density of $n_H = 100$ cm$^{-3}$. The escape fraction of the ionizing radiation from the H II region is assumed to be zero (Byler et al. 2017).

We then measure the three spectral features of interest following the method outlined in Section 2.2 and also compute the actual SFR$_{79}$ for each of the mock SFHs. Note that for the mock galaxies, the EW(H$\alpha$) is measured including only the contribution of nebular emission by newly formed stars. This appears to be different from the measurement of EW(H$\alpha$) in the observation, which also includes the contribution of post-AGB stars. Therefore, we later re-examine our main results, subtracting the contribution of LINERs (Low-Ionization Nuclear Emission Regions; Baldwin et al. 1981; Kewley et al. 2006) to the EW(H$\alpha$) (see Section 2.3.4 and Appendix C).

Compared to the original method of WL20, we here employ an improved set of SFHs. Instead of using the log-normal fits to the SFHs of galaxies from the Illustris simulation from Diemer et al. (2017) as in WL20, we instead start with the SFHs for a wide range of stellar masses constructed from the "observed" evolution of the SFMS (Stark et al. 2013; Speagle et al. 2014; Lilly & Carollo 2016), which are likely to be more realistic. Specifically, for a given galaxy with a given stellar mass at the current epoch, we can move backward to derive the stellar mass at a lookback time of $\Delta t$ based on the current SFMS. In the same way, we can then obtain the stellar masses of this galaxy at a lookback time of $2\Delta t$, $3\Delta t$, etc. based on the SFMS of the corresponding epochs, i.e., the stellar-mass assembly history of this galaxy. This can further be converted to the SFH of this galaxy with the time-evolving SFMS (also see Section 7.3 in Peng et al. 2010).

In addition, since in this work we are interested in galaxies that are potentially undergoing quenching, we also include SFHs in which the SFR is irrevocably declining. A simple quenching model was imposed on half of the SFHs by multiplying by an exponential function:

$$Q(t) = \begin{cases} 1 & t < \tau_S \\ \exp\left(\frac{\tau_S - t}{\tau_Q}\right) & t > \tau_S, \end{cases} \quad (1)$$

where $\tau_S$ represents the starting time of quenching, and $\tau_Q$ is the e-folding time of quenching. A wide range of both $\tau_Q$ and $\tau_S$ is implemented in the model SFHs, i.e., $\tau_Q \in [0.06, 10]$ Gyr, uniformly distributed in logarithmic space, and $\tau_S \in [1.1, 13.6]$ Gyr, uniformly distributed in linear space.

In this simple model, quenching goes on "forever," i.e., a given quenching galaxy will keep decreasing its SFR indefinitely, approaching 0 for $t \to \infty$. We believe this to be unrealistic and show and discuss in Appendix D that it leads to unreasonably low values of SFR$_{79}$ for a large number of galaxies with low H$\alpha$ emission. In the final version of the calibrator that is used in this work, we therefore set a "floor" to the quenching process so that quenching stops once a galaxy's SFR has decreased by 1.3 dex (i.e., a factor of $\approx 20$), at which point the galaxy is well off the SFMS. From then on, the SFH remains "flat," i.e., the "quenched" galaxy keeps forming stars at a low and constant SFR. We discuss the effect of the choice of a prior distribution of quenching SFHs on our results in Section 2.3.5 and Appendix D. We then construct the final set of mock SFHs by adding short-term stochastic variations to all of these smoothly varying SFHs. This follows the same procedure as in WL20, but with a slightly larger amplitude (~0.5 dex in standard deviation) to cover more possibilities. We wish to emphasize that the inclusion of quenching SFHs in our calibrator does not induce any circularity in our approach of trying to find observational evidence for ongoing quenching. The mock SFHs are used to convert an observed set of spectral indices to a value of SFR$_{79}$. In this procedure, which is described in more detail below, we do not keep track of the individual mock SFHs that are associated with a given set of spectral indices but average over all mock SFHs that can reproduce the given spectral features within the observational errors. In this sense, the inclusion of quenching simply adds to the diversity of SFHs in the calibrator and allows us to cover a larger part of the H$\alpha$–H$\delta$ parameter space with our mock galaxies.

Figure 3 shows the distribution of the mock spectra on the log(EW(H$\alpha$))–EW(H$\delta_A$) plane (left panels) and on the log(EW(H$\alpha$))–$D_n$4000 plane (right panels), color-coded by their median log(sSFR$_7$) (top panels), log(sSFR$_9$) (middle panels), and log(SFR$_{79}$) (bottom panels). For comparison, we show the distribution of our SDSS sample galaxies in black solid contours. As can be seen, the range of observed spectra is well covered by the mock spectra, except for the lowest values of EW(H$\delta_A$) and highest values of $D_n$4000. The coverage of those regions would be slightly better, but still incomplete if we adopted the model where quenching goes on "forever," indicating that some galaxies do in fact decrease their SFR by substantially more than 1.3 dex. However, these objects would likely have quenched more than 1 Gyr ago and are therefore out of the scope of this paper (see Section 2.3.3). Further, it should also be noted that the lowest measured values of EW(H$\delta_A$) are likely the result of noise as the distribution of EW(H$\delta_A$) at very low EW(H$\alpha$) ~ 1 Å is dominated by observational noise.

In all panels, the color coding shows a strong and clear gradient. In the top panels, this gradient is almost entirely horizontal while in the middle panels it is mostly vertical, indicating that log(sSFR$_7$) is strongly correlated with log(EW(H$\alpha$)), as expected, while log(sSFR$_9$) is correlated with both EW(H$\delta_A$) and $D_n$4000. The clear gradients in the bottom panels illustrate that log(SFR$_{79}$) is indeed well-determined by the combination of log(EW(H$\alpha$)), EW(H$\delta_A$), and $D_n$4000 over a large volume in parameter space.

Instead of using an analytic formula to calibrate the SFR$_{79}$ as in WL20, we here construct a three-dimensional lookup table based on all the mock spectra. Specifically, for a given combination of EW(H$\alpha$), EW(H$\delta_A$), and $D_n$4000 observed in an SDSS galaxy, together with the corresponding observational uncertainties, we sample 1000 SFHs from the mock catalog whose spectral features constitute a three-dimensional Gaussian distribution in (EW(H$\alpha$), EW(H$\delta_A$), $D_n$4000) parameter space with mean equal to the 3-tuple of the observed spectral features and dispersion in each "dimension" given by the respective observational uncertainty. We then obtain the SFR$_{79}$ as the median value of these 1000 mock SFHs and also get an estimate of the uncertainty in this quantity, $u_{SFR79,tot}$, from the rms scatter of the SFR$_{79}$ in these 1000 sampled SFHs.

We note again that all of the spectra-based measurements only apply to the central region of each galaxy that falls within the 3″ SDSS fiber. We therefore convert those measurements of the SFR to specific SFRs by dividing by the stellar mass in the fiber given in the MPA-JHU catalog.

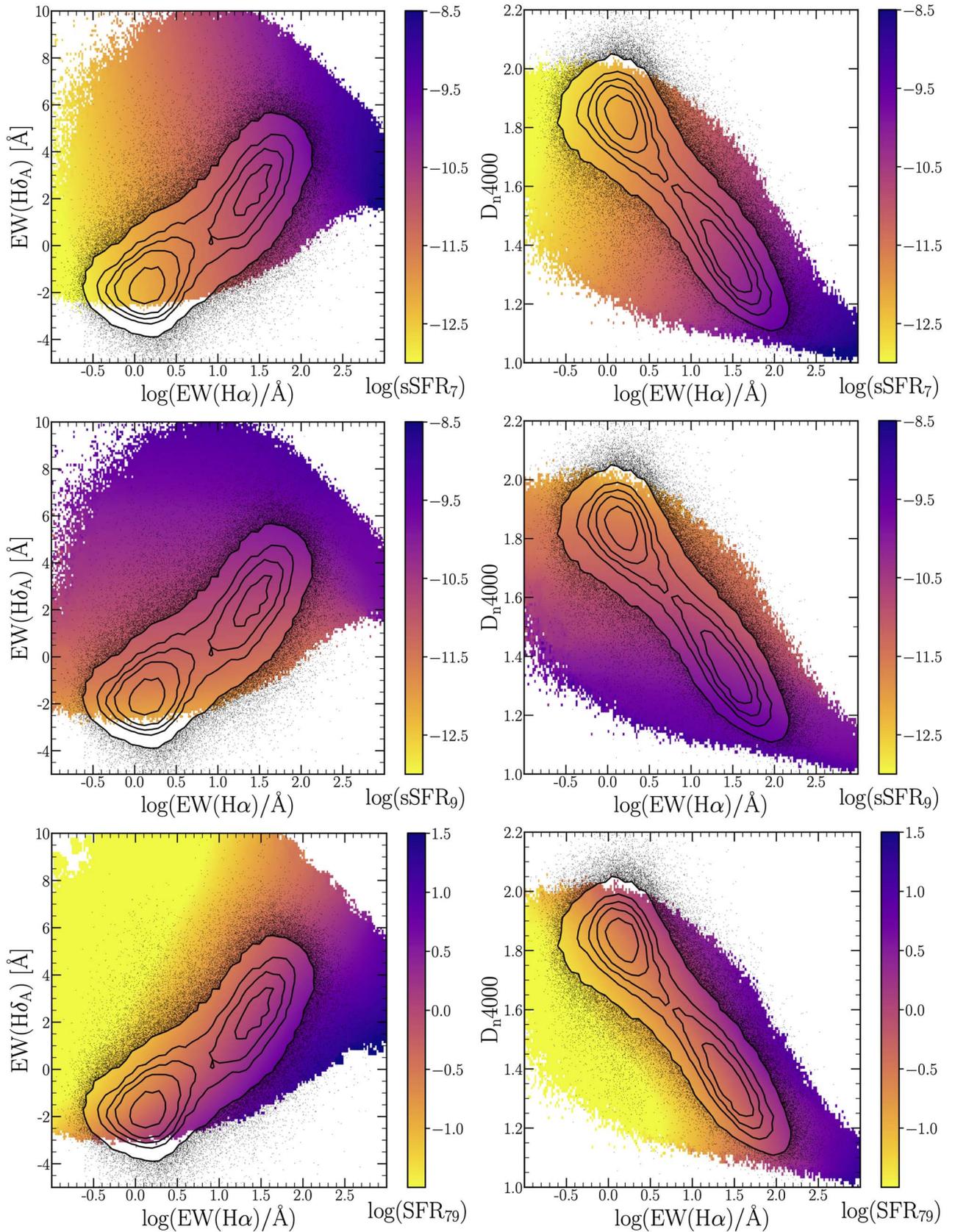

**Figure 3.** Distribution of the ∼2 × 10$^7$ mock spectra used to calibrate the star formation change parameter SFR$_{79}$ on the log(EW(Hα))–EW(H δ$_A$) (left panels) and the log(EW(Hα))–$D_n$4000 (right panels) diagram, color-coded with median log(sSFR$_7$), log(sSFR$_9$), and log(SFR$_{79}$) from top to bottom. For comparison, we show the distribution of our sample galaxies from SDSS DR7 in black solid contours and small dots. The contours include 10%, 30%, 50%, 70%, and 90% of our sample, respectively. The black dots show objects beyond the 90% contour.

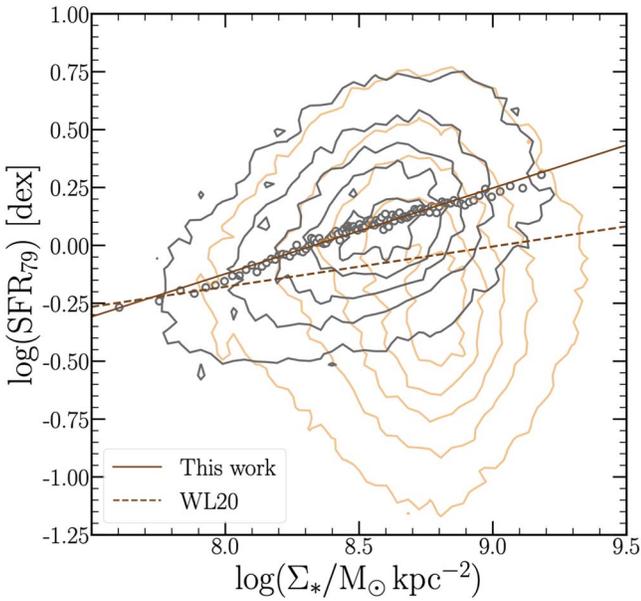

**Figure 4.** Correction of the dependence of log(SFR$_{79}$) on log($\Sigma_*$) based on the galaxies within ±0.3 dex of the fitted SFR$_7$-based SFMS (see Section 3.1). The solid-gray contours show the distribution of these SF galaxies, including 10%, 30%, 50%, 70%, and 90% of them. The median log(SFR$_{79}$) as a function of log($\Sigma_*$) for 500 consecutive objects respectively is shown by the gray circles, and our fit to this median relation by the brown solid line. For comparison, the brown dashed line shows the analogous relation used to correct a similar dependence in WL20. The orange solid contours represent the full sample, analogous to the gray ones.

### 2.3.2. The Ad Hoc Correction of the SFR$_{79}$–$\Sigma_*$ Dependence

By studying the spaxels of spatially resolved Mapping Nearby Galaxies at Apache Point Observatory (MaNGA; Bundy et al. 2015) galaxies, WL20 found that the derived SFR$_{79}$ weakly increases with stellar surface density ($\Sigma_*$), suggesting that SF galaxies have on average a slightly negative SFR$_{79}$ radial gradient. This is not likely to be real, because we would in fact expect the opposite trend in any "inside-out" scenario of galaxy evolution (see the discussion in Section 3.3 of WL20).

We examine the dependence of SFR$_{79}$ on the average $\Sigma_*$ within the fiber for galaxies that are selected to be within ±0.3 dex of the fitted SFR$_7$-based SFMS (see Section 3.1). The resulting contours and median relation are shown in Figure 4. We have corrected the inclination effect when calculating $\Sigma_*$ by multiplying the minor-to-major axis ratio measured in the r-band image.

As can be seen, we find a trend similar to but even stronger than that in WL20. This could be due to the use of the 3″ fiber spectra, meaning that we only investigate the central regions of SDSS galaxies while the MaNGA sample used in WL20 is dominated by the outer regions of galaxies.

As in that previous work, the physical origin of the trend in Figure 4 is not completely understood. It could be due to a dependence on $\Sigma_*$ of the metallicity, of the IMF or of a broadening of the stellar absorption (see also WL20), or some combination of these. Following the same approach as in WL20, we apply an ad hoc correction to the values of SFR$_{79}$, in order to eliminate the dependence of SFR$_{79}$ on $\Sigma_*$. To do this, we first fit a straight line to the median relation of log(SFR$_{79}$) versus log($\Sigma_*$) of the galaxies that are located within 0.3 dex of the fitted SFR$_7$-based SFMS (Section 3.1, shown in light gray in Figure 4). We then use this line to correct for the dependence by assuming that the median log(SFR$_{79}$) of these SFMS galaxies at a given log($\Sigma_*$) equals zero. This assumption is equivalent to the statement that there are as many objects with a recently enhanced SFR as there are objects with a recently suppressed SFR, with respect to the SFR averaged over the previous ~800 Myr. This is a reasonable assumption for galaxies close to the ridge line of the SFMS, as we will further discuss in the context of a possible quenching signature in Section 4.2.1. Moreover, a median log(SFR$_{79}$) significantly different from 0 for these galaxies would indicate a cosmic evolution of the SFMS, which is inconsistent with observations (see, WL20). This ad hoc correction, derived from the restricted set of SF galaxies near the ridge line of the SFMS, is then applied to all of our sample galaxies.

Even though this correction is quite substantial, we emphasize that it will not change any of our basic conclusions since we are mainly interested in the scatter in log(SFR$_{79}$) or its relative values with respect to the values found in typical SFMS galaxies. The effect of this correction on the presented results will be discussed subsequently whenever it is relevant.

### 2.3.3. Uncertainty in SFR$_{79}$

The uncertainty $u_{\mathrm{SFR}_{79},\mathrm{tot}}$ of SFR$_{79}$ (introduced in Section 2.3.1) consists of two parts: (1) the observational uncertainty in the input measurements of the spectral features $u_{\mathrm{SFR}_{79},\mathrm{obs}}$, and (2) the uncertainty that is intrinsic to the calibration of the estimator, $u_{\mathrm{SFR}_{79},\mathrm{cal}}$. The latter is due to the fact that there is not a unique match between SFR$_{79}$ and the three spectral features. Galaxies with the same SFR$_{79}$ can in principle have different spectral features because a range of SFHs can have the same value of SFR$_{79}$. The corollary of this is that a range of different SFR$_{79}$ can produce the same 3-tuple of spectral features.

If our mock SFHs constitute a good representation of the range of SFHs exhibited by galaxies in the real universe, then the $u_{\mathrm{SFR}_{79},\mathrm{tot}}$ will be a reasonable estimate of the real uncertainty. The SFH variation among the mock galaxies is likely, however, by construction, to be larger than that in the real universe. This means that $u_{\mathrm{SFR}_{79},\mathrm{cal}}$ and therefore $u_{\mathrm{SFR}_{79},\mathrm{tot}}$ likely overestimate the true uncertainty of SFR$_{79}$ for any real galaxy. In practice, the real uncertainty in SFR$_{79}$ should lie somewhere between $u_{\mathrm{SFR}_{79},\mathrm{obs}}$ and $u_{\mathrm{SFR}_{79},\mathrm{tot}}$.

Figure 5 shows the rms $u_{\mathrm{SFR}_{79},\mathrm{tot}}$ for our full calibrator as a function of log(EW(H$\alpha$)) as the gray solid line. It is calculated as the scatter of log(SFR$_{79}$) among mock galaxies within grid cells in parameter space whose width is given by ±1$\sigma$, where $\sigma$ represents the median observational uncertainties in each spectral feature. The gray dashed line is obtained by computing the scatter of log(SFR$_{79}$) in very small grid cells in parameter space and therefore provides an estimate of the uncertainty intrinsic to the calibrator, $u_{\mathrm{SFR}_{79},\mathrm{cal}}$. Based on the three observed spectral features and their uncertainties, log(SFR$_{79}$) can be determined with an uncertainty of ~0.15 dex or less for SF galaxies, i.e., those with EW(H$\alpha$) > 7 Å shown here in the blue region (see Section 3.1). For EW(H$\alpha$) < 4 Å (shown in the red region), the total uncertainties in SFR$_{79}$ increase markedly, while the uncertainty intrinsic to the calibrator increases more modestly, indicating that the uncertainty in SFR$_{79}$ below 4 Å is dominated by the observational uncertainty in log(EW(H$\alpha$)), which increases rapidly relative to the measured emission in logarithmic space as we move toward lower EW(H$\alpha$). As can be seen in, e.g., Cid Fernandes et al. (2010), the distribution of log (EW(H$\alpha$)) of the SDSS galaxies clearly shows the well-known bimodality with the peak of the quenched population at around EW(H$\alpha$) ~ 1 Å. The typical observational uncertainty in our measurement of EW(H$\alpha$) around this peak is ~0.6 Å, which is certainly a major contributor to the increase in the uncertainty of SFR$_{79}$ below 4 Å. For EW(H$\alpha$) > 4 Å, however, the measurements of EW(H$\alpha$) are reliable in the sense that above 4 Å, the measured emission is highly unlikely to be just noise.

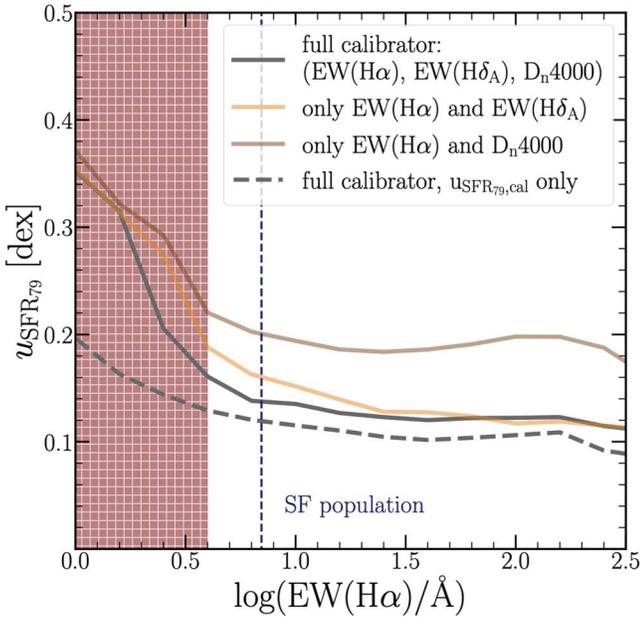

**Figure 5.** Uncertainty of the inferred log(SFR$_{79}$) as a function of log(EW(H$\alpha$)). The solid lines indicate the dispersion in log(SFR$_{79}$) among the mock galaxies within grid cells in parameter space whose size is given by the $\pm 1\sigma$ observational uncertainties in the spectral features as a function of log(EW(H$\alpha$)). This provides an estimate of the $u_{\mathrm{SFR}_{79},\mathrm{tot}}$ introduced in Section 2.3.1. The three lines correspond to the full three-dimensional grid cells (gray line) and two-dimensional grid cells in log(EW(H$\alpha$)) and EW(H$\delta_A$) (light-orange line) and in log(EW(H$\alpha$)) and $D_n$4000 (light-brown line). The gray dashed line indicates the dispersion in much smaller (three-dimensional) grid cells and therefore provides an estimate of the calibrator uncertainty $u_{\mathrm{SFR}_{79},\mathrm{cal}}$ alone. A cut at EW(H$\alpha$) = 7 Å roughly corresponding to the definition of the SF population in Section 3.1 is indicated by the blue dashed line, and the region with EW(H$\alpha$) < 4 Å, which we adopt as a cut on the reliability of the calibration, is shown as the red shaded region.

We have also examined the role of the two parameters EW(H$\delta_A$) and $D_n$4000 in determining SFR$_{79}$ by calculating the $u_{\mathrm{SFR}_{79},\mathrm{tot}}$ in a similar way to that above but based on just a two-dimensional parameter space. The brown and orange solid lines in Figure 5 show the same $u_{\mathrm{SFR}_{79},\mathrm{tot}}$ for a hypothetical calibration of SFR$_{79}$ that is only based on two spectral features. As can be seen, the SFR$_{79}$ can be well determined by EW(H$\alpha$) and EW(H$\delta_A$) for galaxies with EW(H$\alpha$) $\gtrsim$ 10 Å, while $D_n$4000 significantly improves the calibration of SFR$_{79}$ for galaxies with lower EW(H$\alpha$).

If we attempt to calibrate SFR$_{79}$ based on only EW(H$\alpha$) and $D_n$4000, $u_{\mathrm{SFR}_{79},\mathrm{tot}}$ increases by the equivalent of adding ~0.15 dex in quadrature for 4 Å $\lesssim$ EW(H$\alpha$) $\lesssim$ 250 Å and by slightly less (or more) at higher (or lower) values of EW(H$\alpha$). This means that the uncertainty of the calibration roughly doubles for typical SF galaxies if only EW(H$\alpha$) and $D_n$4000 are used. It should be noted that, because in this case the uncertainty mainly comes from the uncertainty intrinsic to the calibrator, more accurate measurements of EW(H$\alpha$) and $D_n$4000 will not significantly help.

Consideration of the uncertainties in SFR$_{79}$ suggests that any sample of sub-SFMS galaxies should be limited to have EW(H$\alpha$) > 4 Å, and we will apply this cut to the analysis in Section 4.

### 2.3.4. Contamination of H$\alpha$ by AGN Emission

The H$\alpha$ emission of a given galaxy is not necessarily due to star formation alone but can be contaminated by emission related to an AGN. Notably, Seyfert galaxies (Maiolino & Rieke 1995; Heckman et al. 1997) and LINERs can contribute to the measured EW(H$\alpha$). However, the fraction of Seyfert galaxies is about ~3%–5% (Pasquali et al. 2009; Wang et al. 2018b), and therefore, their effect on our results is expected to be negligible. As a quick check, we have removed all Seyfert galaxies according to the cuts presented in Kewley et al. (2006) from our sample and found no significant effect on our results.

Even with the cut of EW(H$\alpha$) > 4 Å, one can however not fully ignore the contribution of LINER emission to the EW(H$\alpha$). Belfiore et al. (2016) found that the LINER emission tightly follows the continuum due to the underlying old stellar population (OSP), so it usually contributes only weakly to the equivalent width, i.e., EW(H$\alpha$)$_{\mathrm{LINERs}}$ < 3 Å.

To get an idea of the effect of LINER emission on our results, we may apply an empirical correction to the H$\alpha$ emission based on the classification of our sample galaxies on the Baldwin, Phillips & Terlevich (BPT) diagram (Baldwin et al. 1981; Kauffmann et al. 2003; Kewley et al. 2006). We refer the reader to the details of this correction in Appendix C. We note that this LINER correction moves 15,525 or 12.7% of the galaxies that originally had EW(H$\alpha$) > 4 Å to below this selection boundary.

We note that this correction is applied to all galaxies classified as an AGN on the BPT diagram, including Seyfert galaxies whose potential contribution to the H$\alpha$ emission is therefore at least partly accounted for by the LINER correction.

In Section 4, we will therefore always show, either in the main text or in the Appendix, two versions of our main results, with and without this LINER correction, and we will discuss the implications of possible differences. The results in Section 3 regarding SFMS galaxies are almost entirely unaffected by this correction because the overwhelming bulk of these objects have much stronger H$\alpha$ emission.

### 2.3.5. The Effect of the Quenching SFH Prior

In Appendix D, we examine whether the estimates of SFR$_{79}$ are significantly affected by the choice of the prior in the distribution of quenching SFHs that were used in the construction of the mock spectra. We examine two different representations of the quenching process. We find that the choice of prior indeed has a substantial and systematic effect on the inferred values of SFR$_{79}$ at low EW(H$\alpha$) but that for the sample above, the proposed cut of EW(H$\alpha$) > 4 Å, the effect is negligible.

### 2.3.6. The Effect of Additional Old Stellar Populations on the Estimation of SFR$_{79}$

A final question is whether the addition of a substantial OSP will perturb the estimate of SFR$_{79}$. We here consider an OSP to be one in which all the stars are much older than 1 Gyr and which therefore contribute nothing to SFR$_9$ (or SFR$_7$). A composite system, consisting of a "normal" SF component plus a substantial additional OSP, would have a decreased sSFR$_9$ (and sSFR$_7$) but the value of SFR$_{79}$ should not, at least in principle, be affected. Such a galaxy would not normally be considered to be "quenching" as it would have a more or less constant SFR on gigayear timescales, and in particular, it would have SFR$_{79} \sim 1$.

An important question is whether the addition of such an OSP could nevertheless spuriously bias our estimate of SFR$_{79}$. This is investigated in Appendix E, in which we examine how a system moves on the log(SFR$_{79}$)–$\Delta$log(sSFR$_9$) diagram when we progressively add such an OSP to an otherwise normal SFMS system.

To summarize the result, we first note that for an OSP with an age of 2, 5, or 10 Gyr, a mass roughly 2 to 3 times the integrated mass of the original SF component is required to move a composite object out of the SF population into the GV region. In the case of older OSP ages, 5 Gyr and greater, application of our standard SFR$_{79}$ estimator (correctly) returns the SFR$_{79} \sim 1$ of the SF component. Such a composite system would therefore not mimic a galaxy with a currently significantly declining sSFR.

If the added OSP is younger, i.e., 2–5 Gyr, then the estimator may well return a falsely low value of log(SFR$_{79}$) $\approx 0.4$, suggesting a declining SFR. However, a galaxy that consists of a continually forming population of a certain integrated mass plus an OSP that is 2–3 times as massive but formed only 2–5 Gyr previously would represent a rather odd SFH. We believe that such galaxies will be rare in the universe. Even if the derived SFR$_{79}$ is biased, such objects could anyway more justifiably be considered to be "quenching" since the SFR would have dramatically declined over the last few gigayears.

## 3. The SFMS Population

In this section, we will introduce the SFR–$M_*$ diagrams on two different averaging timescales and provide a definition of the SFMS ridge line as well as of the SF population in Section 3.1. We will then present an overview and a brief discussion of the typical values of SFR$_{79}$ found in different regions of the SFR–$M_*$ diagrams in Section 3.2.

### 3.1. The Definition of the SFMS

After obtaining the SFR$_7$ and SFR$_9$ of each individual galaxy, we first investigate the recent change in the SFR (i.e., the SFR$_{79}$ value) for galaxies in different regions of the SFR–$M_*$ plane. In principle, this directly tells us about the time variation of the SFR of SFMS galaxies.

Figure 6 shows the two SFR–stellar-mass relations based on the SFR averaged over different timescales and color-coded with the median log(SFR$_{79}$). We again emphasize that all the star formation parameters (SFR$_7$, SFR$_9$, SFR$_{79}$) and here also the stellar mass (denoted as $M_{*,\text{fib}}$ in Figure 6) are measured within the SDSS fiber aperture and therefore apply to those regions of the galaxies rather than the whole galaxies. On the left, we plot the entire sample while on the right, we only plot SF galaxies, defined to be objects less than 0.7 dex below the fitted SFR$_7$-based SFMS (see below). This is because the calibration of SFR$_{79}$ (and thus sSFR$_9$) becomes unreliable for low values of EW(H$\alpha$) (or sSFR$_7$), and this can have misleading effects on the color coding on the right of Figure 6.

In both panels, the gray contours indicate the number density of the sample galaxies on the SFR–stellar-mass plane, enclosing 10%, 30%, 50%, 70%, and 90% of the objects. In the left panel of Figure 6, the bimodality of the galaxy distribution is clearly seen. This enables us to separate the main SF population from the quenched population and to define the ridge line of the SFMS in the following way. First, we select by eye a straight line in logarithmic space dividing the two populations. Then, we fit a straight line to the objects above that line, shift it down by 0.6 dex, and use this as the new dividing line. We iterate the above procedure 20 times, which is sufficient to reach convergence.

The resulting SFMS can then be described by the relation

$$\log\left(\frac{\text{SFR}_7}{M_\odot \, \text{yr}^{-1}}\right) = 0.92 \log\left(\frac{M_{*,\text{fib}}}{M_\odot}\right) - 9.39 \quad (2)$$

and is shown as the black solid line in the left panel of Figure 6. In the following, we use this relation as the definition of the ridge line of the SFR$_7$-based SFMS (subsequently referred to as the SFMS$_7$). Based on that, we define the parameter $\Delta$log (sSFR$_7$) to quantify the vertical deviation in dex of log(SFR$_7$) (or thus also log(sSFR$_7$)) from the fitted SFMS$_7$ ridge line at a given stellar mass. This parameter measures the enhancement or suppression of the sSFR$_7$ relative to the SFMS. Several lines of constant $\Delta$log(sSFR$_7$) are drawn on the left panel of Figure 6 for orientation.

We note that the fitted SFMS$_7$ also provides a reasonable fit to the SFR$_9$-based SFMS, as shown in the right panel of Figure 6. Strictly speaking, this is a consequence of the ad hoc correction (see Section 2.3.2) setting log(SFR$_{79}$) to zero, but it also holds well without the correction. This reflects the fact that the mean SFH of individual galaxies on the SFMS has changed only weakly over the last gigayear. WL20 calculate the expected cosmic evolution of the mean log(SFR$_{79}$) to be $\sim -0.025$ dex based on the SFMS from Lilly & Carollo (2016). We consider this to be negligible in the analyses that follow. Therefore, we also use Equation (2) to define the SFR$_9$-based SFMS, denoted as SFMS$_9$ for short. We then define the $\Delta$log(sSFR$_9$) as the vertical deviation in dex of log(SFR$_9$) (or log(sSFR$_9$)) from the SFMS$_9$ at a given stellar mass.

We further define a population of SF galaxies to consist of all galaxies with $\Delta$log(sSFR$_7$) $> -0.7$ dex, i.e., objects above the red dashed line in the left panel of Figure 6. While this definition is somewhat arbitrary, we note that the basic results of this paper are not sensitive to the exact definition of the SF population. We have checked that if we shift the dividing line up by 0.1 dex and define the SF population as $\Delta$log(sSFR$_7$) $> -0.6$ dex, the basic results of this paper do not change. From visual inspection of, e.g., Figure 6, it seems to us that setting the cut significantly higher than 0.6 dex or lower than 0.7 dex below the SFMS ridge line is not reasonable, as it would either cut off a substantial part of the tail of the SF population or start to include a significant number of quenching or quenched galaxies (see the discussion of quenching effects further below). Further, the dispersion in $\Delta$log(sSFR$_7$) of the so-defined SF population is as large as 0.38 dex, including observational noise. Therefore, the cut we apply very roughly corresponds to the $2\sigma$ lower boundary.

The distribution of log(SFR$_{79}$) of the SF population is quite symmetric with a median of $-0.006$, a mean of $-0.0006$, and a dispersion of 0.32 dex, which is relatively large compared to the rms uncertainties in log(SFR$_{79}$), rms($u_{\log(\text{SFR}_{79}),\text{tot}}$)$= 0.21$, and

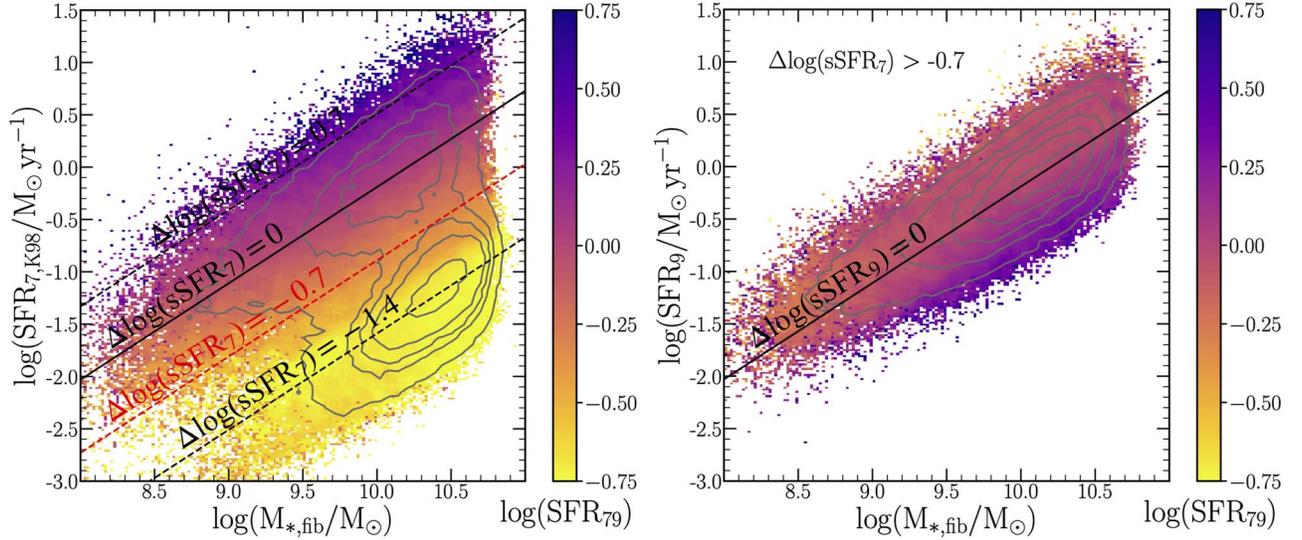

**Figure 6.** log($SFR_7$)–log($M_{*,\mathrm{fib}}$) (left panel) and log($SFR_9$)–log($M_{*,\mathrm{fib}}$) (right panel) relation for the sample galaxies, color-coded with log($SFR_{79}$). We split both planes into a grid of 150 × 225 grid cells and compute the median log($SFR_{79}$) in each grid cell, which we then represent by the color coding after smoothing with a median filter with a window size of 5 × 5 grid cells. The contours including 10%, 30%, 50%, 70%, and 90% of the plotted galaxies are shown as the gray solid lines. The black dashed line labeled "$\Delta$log($sSFR_7$) = 0" in the left panel is the fit to the $SFMS_7$. The quantity $\Delta$log($sSFR_7$) is defined as the vertical deviation of the log($sSFR_7$) from the fitted $SFMS_7$ in dex (see the text for details). Additional lines of constant $\Delta$log($sSFR_7$) are drawn on the left panel for orientation. The line at $\Delta$log($sSFR_7$) = −0.7 is drawn in red, and it represents our definition of the SF population (Section 3.1). While we show the full sample in the left panel, only objects with $\Delta$log($sSFR_7$) > −0.7 (i.e., SF objects) are displayed in the right panel. The line labeled "$\Delta$log($sSFR_9$) = 0" in the right panel is the same line as the fitted $SFMS_7$, which we also use to define an analogous $SFMS_9$ and the deviation from it, $\Delta$log($sSFR_9$).

rms($u_{\log(SFR_{79}),\mathrm{obs}}$) = 0.17 dex (see Section 2.3.3). If we do not apply the ad hoc correction (Section 2.3.2), we find a slightly higher median of 0.07 (mean of 0.08) and a very slightly higher dispersion of 0.33 dex. We will discuss the distribution of $SFR_{79}$ of the SF galaxies in more detail in the context of quenching below.

### 3.2. The $SFR_{79}$ on the SFR–$M_*$ Diagrams

Observationally, the scatter of the SFMS, i.e., the scatter in sSFR at a given stellar mass on the SFMS, varies in the literature between 0.2 and 0.4 dex (e.g., Whitaker et al. 2012; Speagle et al. 2014; Schreiber et al. 2015; Boogaard et al. 2018), depending on the detailed definition of the sample and on how the stellar mass and SFR are measured. There is no evidence for a significant evolution with redshift of the scatter of the SFMS (e.g., Speagle et al. 2014). This stability of the SFMS can be interpreted as the quasi-steady-state interplay between cold gas inflow, star formation, and outflow under a gas-regulator system (e.g., Bouché et al. 2010; Schaye et al. 2010; Davé et al. 2011; Lilly et al. 2013; Wang et al. 2019; Wang & Lilly 2020b).

As discussed further below in Section 4.2.1, the stability of the SFMS requires that the change in the SFR of individual galaxies should not depend on the position of galaxies on the $SFMS_9$ (Wang & Lilly 2020a, 2020b). In other words, the log($SFR_{79}$) should not be correlated with the $\Delta$log($sSFR_9$) for SFMS galaxies. Any such correlation would produce a runaway effect, either broadening or narrowing the width of the SFMS over time. We indeed find that the $\Delta$log($sSFR_9$) and log($SFR_{79}$) for SF galaxies are essentially uncorrelated, as shown in the right panel of Figure 6. The median log($SFR_{79}$) is more or less constant and ≈0 around the $SFMS_9$ and only slightly increases toward the lowest $\Delta$log($sSFR_9$). We have checked that this result does not come from the ad hoc correction applied in Section 2.3.2. The fact that we do not see a correlation is an important consistency check that our analysis is producing reasonable results. As will become clear below, the slight increase at the bottom of the $SFMS_9$ is primarily caused by our selection of objects with $\Delta$log($sSFR_7$) > −0.7 dex, which introduces a bias toward higher log($SFR_{79}$) for the lowest $\Delta$log($sSFR_9$).

It is then easy to see that log($SFR_{79}$) must be correlated with $\Delta$log($sSFR_7$) for these same SF objects, provided there is any time-variability in their sSFR. This is because $SFR_{79}$ tells us whether an object currently has an enhanced or suppressed SFR with respect to its SFR averaged over a longer timescale. The significant correlation that we observe as a smooth color gradient in the left panel of Figure 6 indicates that there is a significant contribution of short-timescale fluctuations in the SFR to the scatter of the $SFMS_7$. We investigate the variability in the SFR of SF galaxies in more detail in Appendix A, where we present a quantitative comparison with the results given in WL20.

### 4. Searching for Signatures of Ongoing Quenching

The goal of this section is to search for spectroscopic signatures of ongoing quenching in galaxies. The SFMS population considered in Section 3 will be overwhelmingly dominated by galaxies that are not quenching. This therefore necessitates the extension of the study into the region below the SFMS, although it will become clear that the analysis must still consider all galaxies.

As discussed above in Section 2.3, a number of considerations force us to limit the analysis to galaxies with EW(H$\alpha$) > 4 Å. We therefore adopt this cut, resulting in a sample of 122,092 galaxies.

We note that below this cut, there is a population of galaxies with low or even vanishing H$\alpha$ emission, which still show EW(H$\delta_A$) $\gtrsim$ 2 Å. Those galaxies can be seen in the upper-left corner of all the left panels in Figure 3, and they represent the well-known E + A or post-starburst galaxies (Dressler & Gunn 1983;

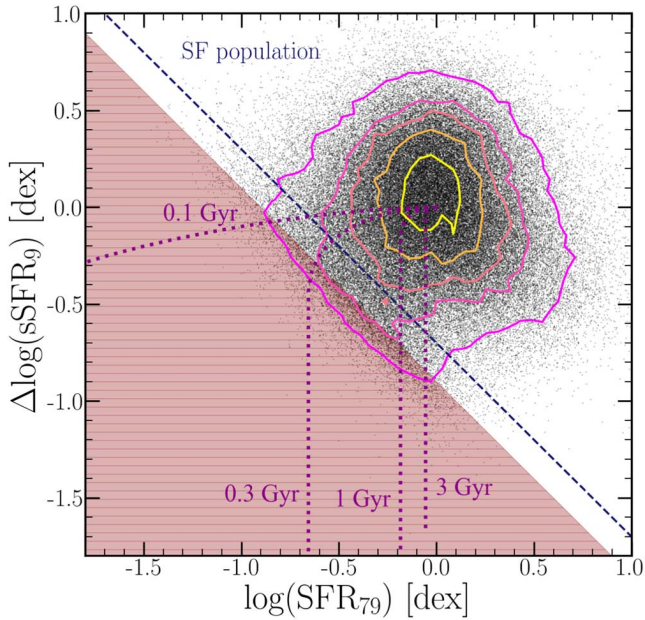

**Figure 7.** log(SFR$_{79}$)–Δlog(sSFR$_9$) diagram for all objects with EW(Hα) > 4 Å. The region corresponding to Δlog(sSFR$_7$) < −0.9, which is roughly equivalent to the sample selection in EW(Hα), is colored in red. Our definition of an SF population used in Section 3 is shown by the diagonal blue dashed line. Possible tracks of a quenching galaxy on the diagram computed as described in the text are shown as magenta dotted lines for four different quenching timescales $\tau_Q$ = 0.1, 0.3, 1, and 3 Gyr.

Zabludoff et al. 1996). Their spectral features require a sudden drop in their SFR, which is converted to a low negative value of log(SFR$_{79}$) in our calibrator as can be seen in Figure 3. While their values of EW(Hα) and EW(Hδ$_A$) can be reproduced by an SFMS SFH undergoing an extreme fluctuation, they are perhaps more plausibly associated with rapid quenching SFHs. If so, the mere existence of those objects itself provides some evidence for ongoing quenching. A more in-depth analysis of those sources is beyond the scope of this work, and since they typically have EW(Hα) < 4 Å, they are subsequently largely ignored.

### 4.1. The log(SFR$_{79}$)–Δlog(sSFR$_9$) Diagram

We now introduce the log(SFR$_{79}$)–Δlog(sSFR$_9$) diagram as a key diagnostic of the change of SFR in galaxies. This is shown for all our sample galaxies with EW(Hα) > 4 Å in Figure 7. The contours enclose 10%, 30%, 50%, 70%, and 90% of the galaxies plotted.

A locus of constant Δlog(sSFR$_7$) is a diagonal line in Figure 7. The region corresponding to Δlog(sSFR$_7$) < −0.9 is shaded in red. This is closely (but not exactly) equivalent to the EW(Hα) < 4 Å cut, according to a linear fit between these two quantities. The blue dashed line indicates the adopted lower limit of Δlog(sSFR$_7$) = −0.7 for the SFMS population examined in Section 3.1. The diagonal strip between this line and the red shaded region may therefore be considered to be a "Green Valley" (GV) sample.

It should be noted that the median log(SFR$_{79}$) and Δlog(sSFR$_9$) of the ridge-line SFMS galaxies are both, largely by construction, zero (see Section 2.3.2).

Note that the large population of already-quenched passive galaxies is expected to lie below the SF population in both Δlog(sSFR$_7$) and Δlog(sSFR$_9$). It is not clear what the SFR$_{79}$ of such galaxies should be, and, as discussed in Section 2.3, we anyway cannot reasonably constrain the SFR properties of this population with our methodology. It is excluded here, but we do show the full original sample in Appendix D, Figure 15, in the context of our discussion of the effect of the assumed prior distribution of SFHs in the calibration of SFR$_{79}$.

It is instructive to consider the possible tracks of a "quenching" galaxy on the log(SFR$_{79}$)–Δlog(sSFR$_9$) diagram. For simplicity, we assume that, prior to the onset of quenching, which occurs at time $\tau_S$, the galaxy has had a more or less flat SFH (e.g., Peng et al. 2010) and thus resided in the middle of the SFMS cloud, i.e., at log(SFR$_{79}$) = Δlog(sSFR$_9$) = 0.

Once it starts quenching at $\tau = \tau_S$, the SFR of this galaxy subsequently declines exponentially with an e-folding timescale of $\tau_Q$ (see Equation (1)). It is then straightforward to compute the track of this galaxy on the log(SFR$_{79}$)–Δlog(sSFR$_9$) diagram. Four quenching tracks with different quenching timescales ($\tau_Q$ = 0.1, 0.3, 1.0, and 3.0 Gyr, respectively) are shown as magenta dotted lines. The quenching galaxies move to lower sSFR$_9$ and lower SFR$_{79}$. If the decline in SFR is a pure exponential then, as shown, the tracks become vertical after 1 Gyr, i.e., as soon as the memory of the pre-quenching state has been lost.

Galaxies leaving the SFMS via quenching will therefore do so via the lower-left quadrant relative to their starting point (assumed here to be the peak of the SFMS population). It can be seen that such galaxies will therefore cross our "GV strip" at a location in log(SFR$_{79}$)–Δlog(sSFR$_9$) that depends on their quenching timescale.

It is evident in Figure 7 that there are indeed galaxies found in the expected location, i.e., in the GV strip just below the SFMS population and with negative values of log(SFR$_{79}$). This is indicated by the distortion of the magenta contours toward the lower left. With the LINER correction, however (Section 2.3.4 and Appendix C), this distortion is largely absent (see Figure 14). Note that the bulk of the objects lies between the tracks corresponding to $\tau_Q$ of 300 Myr and 1 Gyr. Clearly, however, the tracks and the corresponding timescales will depend somewhat on the assumed starting point of the quenching tracks, which may not be the midpoint of the SF population in the real universe, at least not for all galaxies. This is illustrated in Appendix B, where we also show how we derived an analytic relation to compute the quenching tracks.

While the distribution of objects on the log(SFR$_{79}$)–Δlog(sSFR$_9$) diagram is at least consistent with the existence of a subset of galaxies quenching along such tracks, the identification of these objects as "quenching" is by no means trivial, as we will discuss below.

One could imagine that the true two-dimensional distribution of genuine SFMS objects (i.e., of a stable SFMS population without any quenching) in Figure 7 could be symmetrical in both log(SFR$_{79}$) and sSFR$_9$. It is then clear that if we slice off a "GV population" (defined from sSFR$_7$) lying along a diagonal strip of the diagram, some of the objects with low SFR$_{79}$ could simply be the SFMS counterparts of other SFMS galaxies with the same sSFR$_9$ but higher SFR$_{79}$.

This makes clear that the overall distribution of SFR$_{79}$ of galaxies in the entire diagram, as well as how this may be modified by observational uncertainties, must be considered before any conclusions about the presence of a quenching population of galaxies can be drawn. An important conclusion is that it may be very misleading to look only at galaxies in the (SFR$_7$-defined) GV. We address this issue in the following sections of the paper.

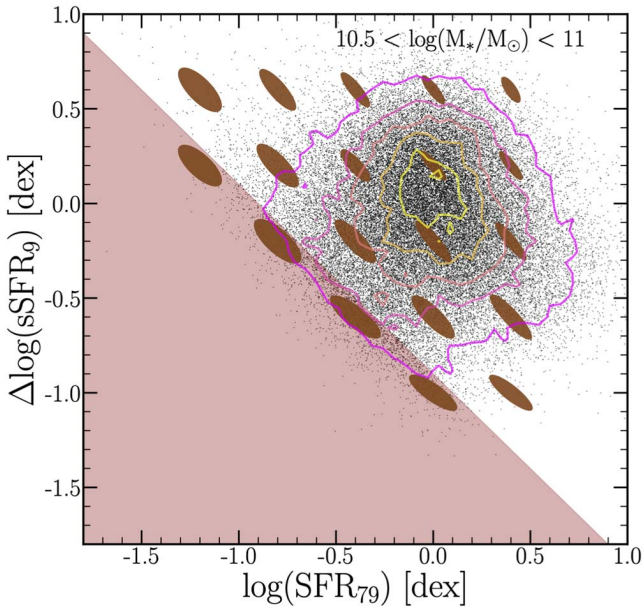

**Figure 8.** log(SFR$_{79}$)–Δlog(sSFR$_9$) diagram in the mass bin $10.5 < \log(M_*/M_\odot) < 11$ with overplotted error ellipses.

### 4.1.1. Effects of Observational Noise on the log(SFR$_{79}$)–Δlog(sSFR$_9$) Diagram

To better understand the effect of noise, both arising from the observational measurements and from the uncertainties inherent in the SFR$_{79}$ calibrator (see Section 2.3.3), on the distribution of objects in the log(SFR$_{79}$)–Δlog(sSFR$_9$) diagram, we show in Figure 8 representative error ellipses across this diagram. These are shown for illustration for galaxies in the mass bin $10.5 < \log(M_*/M_\odot) < 11$.

As discussed earlier, the uncertainty in SFR$_{79}$ mainly comes from the observational uncertainty in EW(H$\delta_A$). Since sSFR$_9$ is derived from SFR$_{79}$, the uncertainty in Δlog(sSFR$_9$) is highly correlated with the uncertainty in log(SFR$_{79}$), as reflected by the diagonal orientation of the error ellipses. As an aside, this indicates that our sample selections in Δlog(sSFR$_7$) and EW(H$\alpha$) should be relatively insensitive to noise.

At fixed Δlog(sSFR$_7$) (i.e., along a diagonal locus in the diagram), there is therefore a substantial contribution of noise to the distribution of log(SFR$_{79}$) (and thus also Δlog(sSFR$_9$)). In particular, at the lower boundary of the sample plotted in the figure (i.e., around Δlog(sSFR$_7$) = −0.9), the width of the distribution of log(SFR$_{79}$) is likely to be dominated by observational noise.

Furthermore, the orientation of the error ellipses tells us that noise will tend to disperse the main peak of SFMS objects in a diagonal direction, scattering them toward the upper left and lower right. One important consequence of this is that noise will amplify the lower-right extension of the SFMS cloud that balances the lower-left extension, which lies in the same location as any quenching galaxies.

This makes clear that in such a diagram, the galaxies in the GV strip cannot be considered in isolation but only in the context and modeling of the entire SFMS population above it, and it is to this that we now turn.

### 4.2. The Expected Signature of a Quenching Population

We here try to derive quantitative expectations of quenching signals. We will start in Section 4.2.1 by deriving and discussing the general stationarity criterion for the SFMS population and then show how the addition of a subset of quenching galaxies will perturb this using the analytic scheme outlined in Peng et al. (2010) and Peng et al. (2012) to estimate the current quenching rate of galaxies and thus the expected number of quenching galaxies as a function of mass. We then examine the effect of the H$\alpha$ selection and of observational scatter before comparing this expectation to the data.

#### 4.2.1. The Stationarity Criterion for a Stable SFMS Population

It is clear from Figure 7 that the low values of SFR$_{79}$ seen in the GV strip are not more extreme than those exhibited by galaxies on the ridge line of the SFMS at Δlog(sSFR$_9$) ∼ 0. In a formal sense, it is evidently therefore not possible (at least with the current methodology) to separate quenching galaxies from SFMS galaxies only on the basis of their individual past SFHs. Such an analysis must be statistical.

Not least, the cut at Δlog(sSFR$_7$) = −0.7 used to define the SFMS population in Section 3.1 is somewhat arbitrary and does not mean that there are no SFMS objects whatsoever at Δlog(sSFR$_7$) < −0.7, i.e., in our GV strip. Given the relatively small number density of galaxies right below the SF population (see Figure 7), even a small tail of SFMS galaxies at Δlog(sSFR$_7$) < −0.7 could dominate the galaxy population in this part of the diagram (see the discussion in Section 4.4).

We therefore need to examine whether the overall distribution of galaxies on the log(SFR$_{79}$)–Δlog(sSFR$_9$) diagram is consistent with an SFMS population alone or whether there is evidence for an additional quenching population. We therefore need to first establish the expected signature of ongoing quenching, and then search for it in the data.

We therefore examine the stationarity criterion for a stable SFMS population. Stationarity means that the shape of the SFMS$_9$ distribution should be constant over time, i.e., $dN(\log(sSFR_9))/dt = 0$, where N represents the number distribution of galaxies in log(sSFR$_9$). Given that the value of SFR$_{79}$ tells us how fast a galaxy is changing its sSFR$_9$, it is straightforward to prove that this implies, at any and all sSFR$_9$, that the distribution of SFR$_{79}$ must satisfy the following condition:

$$\text{mean}(SFR_{79}) - 1 = 0. \quad (3)$$

This is equivalent to saying that the net flux of galaxies through a given value of sSFR$_9$ must vanish for any stable population since any net flux would imply that the distribution of sSFR$_9$ of SFMS galaxies would be changing with time.

#### 4.2.2. Expected Effect of a Quenching Population on Δmean(log(SFR$_{79}$))

As a corollary of the stationarity criterion, the systematic decrease in SFR for a set of galaxies that are quenching implies a mean(SFR$_{79}$) < 1 for those galaxies and a net flux toward lower sSFR$_9$. The presence of a set of quenching galaxies will therefore perturb the mean(SFR$_{79}$) of the entire population toward values <1.

The downward flux of quenching galaxies should be essentially independent of Δlog(sSFR$_9$) (in the interval between the initial and final levels of Δlog(sSFR$_9$)) and also independent of the quenching timescale $\tau_Q$. It should be given only by the quenching rate of galaxies (i.e., the probability that a given SFMS galaxy quenches per unit time). The perturbation of the mean(SFR$_{79}$) of the entire population will however increase to lower sSFR$_9$ because of the decreasing number of SFMS galaxies relative to the quenching population. The signature of any quenching subpopulation in the log(SFR$_{79}$)–$\Delta$log(sSFR$_9$) diagram will therefore be a progressive shift of the mean SFR$_{79}$ below unity as we go down in SFR$_9$ toward the lower end of the SFMS.

In practice, this signature of quenching will inevitably be countered by two large observational effects: (i) the applied (and required) selection of objects with EW(H$\alpha$) > 4 Å introduces a strong bias toward higher SFR$_{79}$ by progressively removing all objects with the lowest SFR$_{79}$ at low sSFR$_9$, as easily seen from the red shaded region in Figure 7 and (ii) the effect of observational noise (both observational in the spectroscopic measurements and from the calibration of SFR$_{79}$). This observational noise has two distinct effects that are both important here.

First, the error ellipses in Figure 7 are diagonal, i.e., lie parallel to lines of constant $\Delta$log(sSFR$_7$). Below the peak of the SFMS (i.e., $\Delta$log(sSFR$_9$) < 0) this will always bias the mean SFR$_{79}$ to higher values, since at a given $\Delta$log(sSFR$_9$) more objects have been scattered from above to higher SFR$_{79}$ than have been scattered from below to lower SFR$_{79}$. The converse is true above the peak.

Second, the observational noise is roughly symmetric in log(SFR$_{79}$). The noise of this form will always bias the mean (linear) SFR$_{79}$ toward higher values. In fact, this logarithmic noise dominates the distribution of log(SFR$_{79}$) at low sSFR$_7$. This means that the best estimate of the underlying mean (SFR$_{79}$) is actually the anti-log of the mean(log(SFR$_{79}$)). We should therefore look for deviations from mean(log(SFR$_{79}$)) = 0.

As an aside, it may be noted that the ad hoc correction to the values of SFR$_{79}$ in Section 2.3.2 effectively normalizes the values of log(SFR$_{79}$) to the values found $\pm$0.3 dex around the fitted ridge line of the SFMS$_7$. It thus effectively forces the stationarity criterion to hold for galaxies within $\pm$0.3 dex of the SFMS$_7$ ridge line. This is a reasonable requirement for the current purposes.

In fact, we will define below an empirical observed midpoint of the SFMS population in each mass bin and compute a $\Delta$mean (log(SFR$_{79}$)), relative to the objects just above the SFMS$_9$ ridge line, as a function of $\Delta$log(sSFR$_9$). The following results should therefore be largely independent of the ad hoc correction.

We may proceed to estimate the expected size of the quenching signal as follows. Peng et al. (2010) and Peng et al. (2012) developed an analytic framework for quenching based on a continuity approach to galaxy evolution. This yields the required quenching rates, i.e., the probability that a given SF galaxy quenches per unit time or, equivalently, the number of galaxies that quench in unit time if we multiply by the number of SF galaxies. It is convenient to follow the distinction between so-called "mass-quenching" and "environment-" or "satellite-quenching" introduced by Peng et al. (2010) and Peng et al. (2012).

The strongly mass-dependent mass-quenching rate $\eta_m$ per galaxy follows directly from the negligible cosmic evolution that is seen in the value of the Schechter parameter $M^*$ at $z \lesssim 2$ (Ilbert et al. 2010) and is given (Peng et al. 2010) by

$$\eta_m(m) = \text{SFR}/M^* = \text{sSFR}_{\text{MS}}(M^*) \times (m/M^*)^{1+\beta}, \quad (4)$$

where sSFR$_{\text{MS}}(M^*)$ is the sSFR of the SFMS at the Schechter mass $M^*$ (taken from Peng et al. (2010)) and $\beta = -0.08$ is the logarithmic slope of the SFMS in terms of sSFR (see Section 3.1). We here adopt $M^* = 10^{11} M_\odot$.

To quantify the rate of (mass-independent) satellite quenching, we may assume for simplicity that the satellite fraction of galaxies is constant with time and that the satellite-quenching efficiency $\epsilon_{\text{sat}}$, is also constant, and that it is independent of satellite mass, $\epsilon_{\text{sat}} \sim 0.5$ (Peng et al. 2012). A straightforward calculation then leads to the required rate of (mass-independent) environmental quenching, $\eta_\rho$, averaged across the overall SFMS population of centrals and satellites:

$$\eta_\rho = -(1 + \alpha_s + \beta)\epsilon_{\text{sat}} \times \text{sSFR}_{\text{MS}}(M^*), \quad (5)$$

where $\alpha_s = -1.3$ is the power-law slope of the Schechter mass function of SF galaxies. The total quenching rate of the galaxy population is then given by the sum of rates of the two quenching channels,

$$\eta_{\text{tot}}(m) = \eta_m(m) + \eta_\rho. \quad (6)$$

With these expected quenching rates, it is then possible to calculate the predicted perturbation of the SFMS stationarity criterion due to the presence of a population of quenching galaxies. Specifically, we model the SF population of galaxies as a 2D Gaussian distribution centered at (0, 0) on the log(SFR$_{79}$)–$\Delta$log(sSFR$_9$) diagram. In each mass bin considered below, we infer the dispersion of that Gaussian distribution from the SF population defined in Section 3.1, with the estimated observational noise subtracted in quadrature. Further, we use the mass distribution (within each bin) of the sample galaxies to infer the overall mass-quenching rate, and also the satellite fraction (from the Yang et al. 2007 group catalog) so as to compute the satellite-quenching rate according to Equation (6). We multiply the number of mock SF galaxies by a factor of 100 with respect to the number of sample galaxies to largely eliminate stochastic variations in the results.

Using Equation (6), we can then calculate the expected number of quenching galaxies. We assume that the quenching rate is constant in time over the timescales of interest and, as assumed previously, that the corresponding galaxies are quenching with an exponential timescale $\tau_Q$ and, for illustrative purposes, that they start quenching from the midpoint of the SFMS. This then gives us the distribution of both SF and quenching galaxies on the log(SFR$_{79}$)–$\Delta$log(sSFR$_9$) diagram. We can then optionally apply a selection to mimic the cut at EW(H$\alpha$) of 4 Å that was applied to the real data and can also convolve the distribution on the log(SFR$_{79}$)–$\Delta$log(sSFR$_9$) plane with typical observational noise so as to compare more directly to the real data.

The results of this exercise are shown in Figure 9. As noted above, we measure the mean(log(SFR$_{79}$)) in each bin relative to the mean found in the bin $0 < \Delta \log(\text{sSFR}_9) < 0.2$, for which the mean log(SFR$_{79}$) should be close to 0, since we do not expect a significant fraction of quenching galaxies above the SFMS$_9$. Moreover, we are primarily interested in the differential trend of this normalized $\Delta$mean(log(SFR$_{79}$)) with $\Delta$log(sSFR$_9$), rather than the absolute values.

In Figure 9 we show in the left panels this $\Delta$mean(log(SFR$_{79}$)) as a function of $\Delta$log(sSFR$_9$) for different $\tau_Q$ in a representative mass bin ($10.5 < \log(M_*/M_\odot) < 11$), and in the right panels for a range of masses (the four mass bins used before) and thus quenching rates for a representative $\tau_Q$ of 500 Myr. In the top panels, we show the expected effect for the idealized case without any selection effects and without noise; in the middle panels, we remove all objects with $\Delta\log(\mathrm{sSFR}_9) < -0.9$, which is roughly equivalent to EW(H $\alpha$) < 4 Å; and in the bottom panels, we also add typical observational noise to the star formation parameters.

If there is no noise (i.e., in the top and middle panels), there is by construction zero $\Delta\mathrm{mean}(\log(\mathrm{SFR}_{79}))$ at $\Delta\log(\mathrm{sSFR}_9) > 0$ (since there are no quenching galaxies above the SFMS$_9$), but as we progress to lower sSFR$_9$, there is the expected perturbation toward negative $\log(\mathrm{SFR}_{79})$ for most mass bins and quenching timescales. In the top-left panel, it can be seen that the different curves are largely independent of $\tau_Q$ just below $\Delta\log(\mathrm{sSFR}_9) = 0$ (except for the very rapidly quenching red curve) but then, they progressively diverge toward lower sSFR$_9$. The first effect is because our $\Delta\mathrm{mean}(\log(\mathrm{SFR}_{79}))$ is effectively measuring the flux of quenching objects: Galaxies that are quenching faster (i.e., with shorter $\tau_Q$) will have a more extreme SFR$_{79}$ but there will be fewer of them in an interval of $\Delta\log(\mathrm{sSFR}_9)$ because they are changing their sSFR$_9$ faster. The net effect is therefore independent of $\tau_Q$. This is however only true as long as the underlying stationary SF population is dominant in terms of numbers. As we move toward lower sSFR$_9$, the relative number of quenching galaxies increases and ultimately the $\Delta\mathrm{mean}(\log(\mathrm{SFR}_{79}))$ converges to a value that is uniquely determined by the corresponding $\tau_Q$. This one-to-one match between SFR$_{79}$ and $\tau_Q$ at sufficiently low sSFR$_9$ can already be inferred from the quenching tracks in Figure 7 that become vertically parallel sufficiently far away from the SFMS$_9$.

The difference between different mass bins in the top-right panel is a simple consequence of the increased mass-quenching rate at higher masses, which dominates the total quenching rate at the masses considered here (Equation (6)).

When the cut in sSFR$_7$ is introduced (the middle panels), the quenching signature is weakened substantially as it is overwhelmed by the bias toward higher SFR$_{79}$ with decreasing sSFR$_9$, as can be easily seen in, e.g., Figure 7. Furthermore, this introduces differentiation between different $\tau_Q$, essentially inverting the previous dependence of the signature on $\tau_Q$. At a given low sSFR$_9$, galaxies that are quenching on a shorter timescale will have a lower sSFR$_7$ and are therefore more quickly affected by the cut in sSFR$_7$. Therefore, for the shortest $\tau_Q$ considered (100 Myr), there is hardly any quenching signature left after applying the cut at $\Delta\log(\mathrm{sSFR}_7) = -0.9$, while for the longest timescale (1.3 Gyr), we would still expect $\Delta\mathrm{mean}(\log(\mathrm{SFR}_{79})) \approx -0.1$ at $\Delta\log(\mathrm{sSFR}_9) \approx -0.7$. For $\tau_Q = 500$ Myr, a signature is only seen for $M_* \gtrsim 10^{10.5} M_\odot$. At lower masses, the quenching rate is too small to produce a measurable signature for this quenching timescale.

If we then add typical observational noise, adopting our minimum (i.e., pure observational) noise estimates (see Section 2.3.3), we find a net trend that is actually opposite to the expected quenching signature, i.e., $\Delta\mathrm{mean}(\log(\mathrm{SFR}_{79}))$ is now increasing toward lower sSFR$_9$ over the entire range considered. As discussed in Section 4.2.2 above, this is due to the fact that the noise in the $\log(\mathrm{SFR}_{79})$–$\Delta\log(\mathrm{sSFR}_9)$ diagram scatters objects along lines of constant sSFR$_7$ (see Figure 7). Since the number of objects decreases as a function of sSFR$_9$ below the peak of the SFMS, this implies that at fixed $\Delta\log(\mathrm{sSFR}_9) < 0$, the number of objects scattering from higher to lower sSFR$_9$ and therefore from lower to higher SFR$_{79}$ is larger than vice versa. Further, this effect gets stronger toward lower sSFR$_9$ as the observational noise increases toward lower sSFR$_9$ (and thus lower sSFR$_7$ on average). Despite this reversal in overall slope, there are still some differential effects with $\tau_Q$ and/or mass and the quenching rate that may be searched for in actual data, especially for different $\tau_Q$. We will compare this to the data in the next section.

### 4.3. Comparison to the Data

In order to compare the expected $\Delta\mathrm{mean}(\log(\mathrm{SFR}_{79}))$ shown in Figure 9 to the real data, we will only consider objects with $\Delta\log(\mathrm{sSFR}_7) > -0.9$ in the data. This is roughly but not entirely equivalent to the previously applied cut at EW(H $\alpha$) = 4 Å (as can be seen from the red shaded region in Figure 7). This additional cut leaves us with 114,496 objects.

We show the mean value of $\log(\mathrm{SFR}_{79})$ relative to the mean found in the bin $0 < \Delta\log(\mathrm{sSFR}_9) < 0.2$ for each of our four mass bins in Figure 10 with and without (light lines) the LINER correction (see Section 2.3.4 and Appendix C).

As expected from the bottom panels in Figure 9, we find that the $\Delta\mathrm{mean}(\log(\mathrm{SFR}_{79}))$ increases to positive values with decreasing sSFR$_9$. This increase is qualitatively consistent with the expected effect from Figure 9 once the effect of observational scatter is taken into account (lowest panels). The trend is less pronounced for higher masses in the data, which is also qualitatively consistent with the trend with mass expected from Figure 9 (bottom-right panel).

However, the trend with mass could also possibly reflect a mass-dependent $\tau_Q$ (Figure 9, bottom-left panel), which would then point to longer timescales for higher masses, in tension with results from, e.g., Hahn et al. (2017), who find shorter $\tau_Q$ for more massive central galaxies (see more discussion in Section 4.4). However, disentangling these effects is clearly not possible with these data, given the effects of observational noise.

Not least, the fact that the trend of $\Delta\mathrm{mean}(\log(\mathrm{SFR}_{79}))$ with $\Delta\log(\mathrm{sSFR}_9)$ levels off at $\Delta\log(\mathrm{sSFR}_9) > 0$ in the data, but should continue from our modeled prediction (Figure 9), also suggests caution in any quantitative interpretation of trends in the data. This may indicate that the underlying true SF population is not a Gaussian distribution in logarithmic space as we have assumed in the previous section and/or that we are overestimating the noise in that regime.

In summary, the trends we observe in $\Delta\mathrm{mean}(\log(\mathrm{SFR}_{79}))$ with $\Delta\log(\mathrm{sSFR}_9)$ are at first sight the opposite of the searched-for quenching signal. However, the observed trends are qualitatively consistent with expectations once the substantial effects of the required cut in sSFR$_7$ (or in EW(H$\alpha$)) and of observational scatter are taken into account. We can only conclude that the observational SDSS data is *consistent* with the presence of a quenching population of the expected strength but does not *demand* it, given the selection effect and the noise.

### 4.4. Estimating Quenching Timescales

If we assume that there is indeed ongoing quenching in the galaxy population, can we get even a rough estimate of the relevant quenching timescales $\tau_Q$ from SFR$_{79}$? We have already seen that while, in principle, this could be done based on examining the trend of $\Delta\mathrm{mean}(\log(\mathrm{SFR}_{79}))$ with $\Delta\log(\mathrm{sSFR}_9)$ (Section 4.2.2), this is in practice not possible given the effect of the required selection cut in sSFR$_7$ and the effect of the significant noise in the SFR$_{79}$ estimates of galaxies.

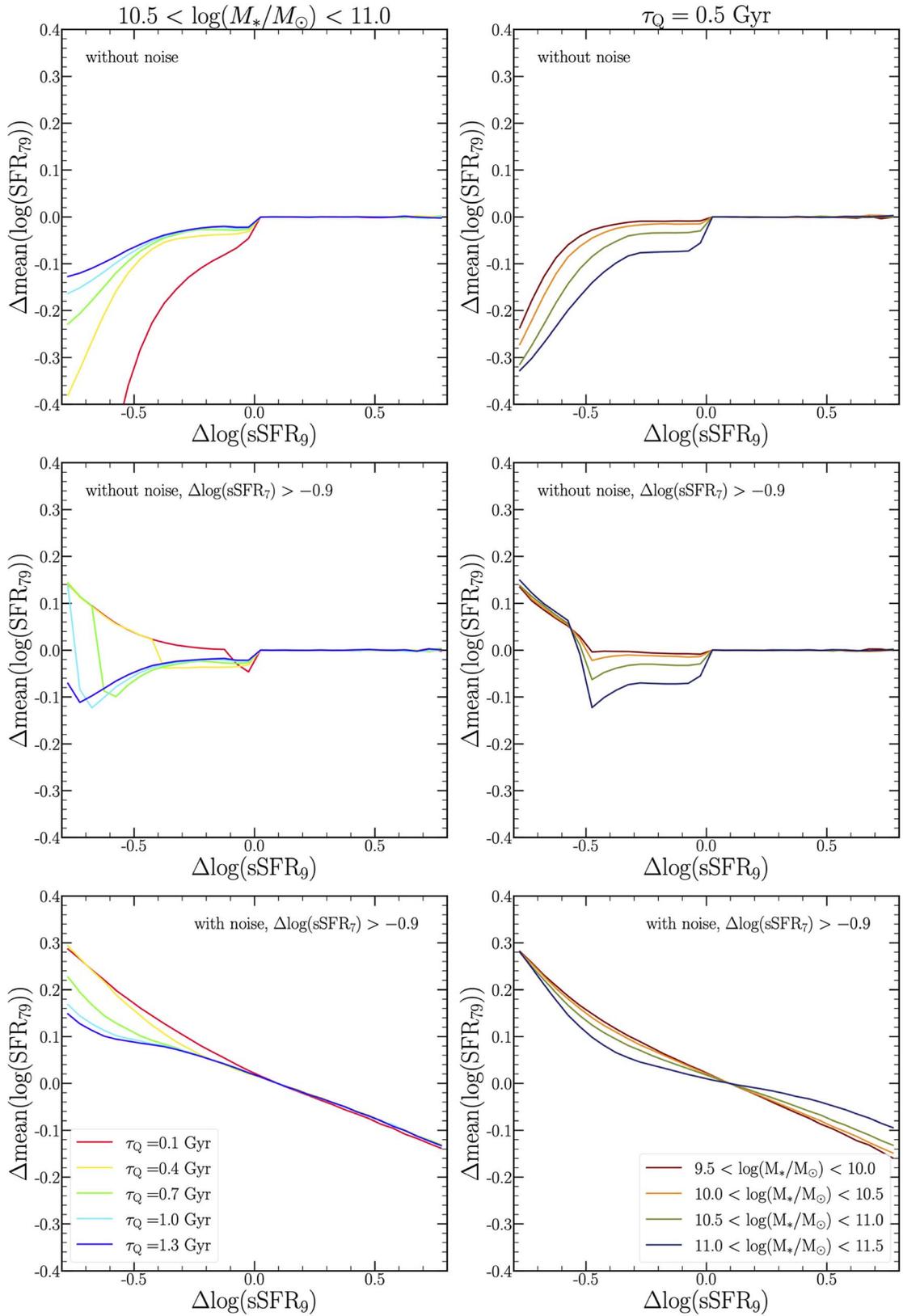

**Figure 9.** Expected effect of the presence of a systematically quenching population on the normalized $\Delta$mean(log(SFR$_{79}$)) as a function of $\Delta$log(sSFR$_9$), as calculated using the quenching framework in Peng et al. (2010) and Peng et al. (2012; see the text for details). On the left, this is shown for a fixed mass bin ($10.5 < \log(M_*/M_\odot) < 11$) and various different quenching timescales $\tau_Q$ and, on the right, for a fixed $\tau_Q = 500$ Myr but the different mass bins used previously (see Figure 1). From top to bottom, the diagrams show the idealized case without noise or selection (top panels), the effect of removing objects with $\Delta$log(sSFR$_7$) < −0.9, roughly corresponding to our selection of objects with EW(H$\alpha$) > 4 Å in the data (middle panels), and finally the effect of adding typical observational noise to each object so as to enable a more realistic comparison to the observational data (bottom panels).

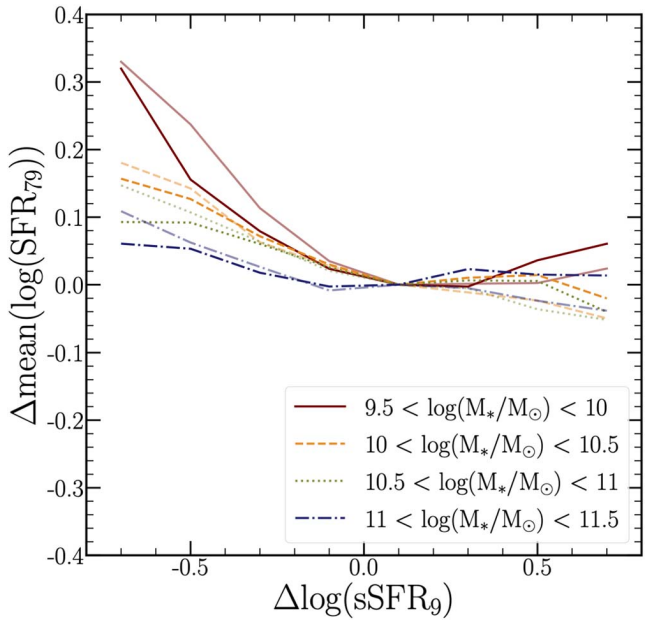

**Figure 10.** $\Delta$mean(log(SFR$_{79}$)), i.e., the mean(log(SFR$_{79}$)) relative to its value in the range $0 < \Delta$ log(sSFR$_9$) $< 0.2$ (see the text for details) in bins of $\Delta$log(sSFR$_9$) that are 0.2 dex wide. The light lines show the analogous quantity derived after the application of the LINER correction (Section 2.3.4 and Appendix C).

If we could isolate a representative "quenching population," we could use SFR$_{79}$ directly to estimate the quenching timescale $\tau_Q$ using the tracks in Figure 7. One possibility could be to use the galaxies in the GV located in the range of $-0.9 < \Delta$ log(sSFR$_7$) $< -0.7$ (corresponding to 4 Å $\lesssim$ EW(H$\alpha$) $\lesssim$ 7 Å), i.e., lying just below the SF population defined above this limit (see Figure 7) and located at the minimum in the number density distribution of log(EW(H$\alpha$)). Such a substantially sub-SFMS population could conventionally be viewed as "quenching," although we will see that this is unlikely to be valid.

Looking at these GV galaxies, the median log(SFR$_{79}$) is found to be $-0.38$, which translates to a $\tau_Q$ of $\sim$500 Myr assuming that the "typical" galaxy follows the quenching tracks in Figure 7 (see also Appendix B for details). Those tracks assume that the galaxies start somewhere close to the ridge of the SFMS population. If quenching instead starts from the top (or bottom) envelope of the SFMS, then the inferred quenching timescales would become slightly longer (or shorter), depending on the detailed assumptions (see Appendix B for a more quantitative discussion of this effect). Furthermore, this direct conversion only provides an estimate of the quenching timescale of the galaxies that are currently seen to be in the GV, i.e., within a narrow range of $\Delta$log(sSFR$_7$), which will be biased against short quenching timescales since these objects will spend little time crossing the SFR$_7$-defined GV. This could in principle be corrected by weighting an observed distribution of $\tau_Q$ by $1/\tau_Q$, but this makes no sense in the present study because we can hardly infer anything about the underlying intrinsic distribution of SFR$_{79}$.

If we have an a priori estimate of the quenching rate of galaxies, following Peng et al. (2010) and Peng et al. (2012; Equation (6)), then we can estimate the number of "quenching galaxies" that should lie in the GV, which also will depend on the typical quenching timescale because this will determine how quickly the galaxies are moving (declining) in SFR$_7$. This second approach is similar to that in, e.g., Wetzel et al. (2013) and Hahn et al. (2017).

The total number of objects in our GV sample is 11,799, corresponding to 10% of the entire sample of galaxies with $\Delta$log(sSFR$_7$) $> -0.9$. Given the quenching rates computed in Section 4.2.2 and using the $\tau_Q \sim 500$ Myr from the previous paragraph, we find that the expected number of "quenching galaxies" that should be seen within our GV is about a factor of 4.5 lower than this actually observed number. This discrepancy decreases to a factor of about 3 when we apply the LINER correction, which hardly affects the median log(SFR$_{79}$) of GV galaxies but does significantly reduce the number of objects in the GV strip relative to the SF population.

This discrepancy in number should not be surprising since we have already argued that the GV population may contain a significant number of SFMS galaxies, i.e., galaxies that should be considered to be members of the "stationary" SFMS population (see the discussion in Section 4.2.1). This is already evident in the $\Delta$log(SFR$_{79}$)–$\Delta$log(sSFR$_9$) plots (e.g., Figure 7). The GV population (defined in terms of SFR$_7$) almost certainly contains a substantial number of low-SFR$_7$ SFMS galaxies, which are the counterparts of the large number of galaxies with the same long-term SFR$_9$ but much higher SFR$_{79}$, and thus higher SFR$_7$, that are visible in that diagram, remembering that any stable SFMS population will satisfy the stationarity criterion of Equation (3).

Another easy way of seeing this is if we make the simplifying assumption that the population of SF galaxies follows a Gaussian distribution in $\Delta$log(sSFR$_7$) centered at 0 and with a dispersion given by the measured dispersion in our SF sample, which is 0.38 dex, including observational noise, 2.4% of the SF population will lie in our GV sample. This means that SF galaxies in the tail of the distribution can account for 20%–30% of the GV population even in this highly simplified scenario. In reality, the distribution of $\Delta$log(sSFR$_7$) is skewed toward negative values, which a priori may or may not be due to quenching effects (see our discussion of the stationarity criterion in Section 4.2.1), potentially increasing the contamination of the GV by SF galaxies.

The conclusion that even 0.8 dex below the ridge line of the SFMS the SFR$_7$-defined GV population may still be dominated by galaxies that should be regarded as being part of the "stationary" SFMS population has important implications for any study that tries to use such GV galaxies as "quenching galaxies."

We note however that the GV sample we examine here is quite different from prevailing definitions in the literature based on photometric colors or the SFR–$M_*$ diagram as they will typically include a broader range of galaxy properties and in particular of sSFR compared to our only 0.2 dex wide "GV strip." Any extension of the GV toward lower sSFR will mitigate the effect of contamination by SF objects.

The likely significant contamination of our GV also implies that the direct estimate of $\tau_Q \sim 500$ Myr based on the median log(SFR$_{79}$) of GV galaxies should be used with caution, as the quenching population could be a minority within that GV. Against this, the distribution of log(SFR$_{79}$) at low $\Delta$log(sSFR$_7$) is dominated by the symmetric Gaussian noise in the data, and the underlying intrinsic distribution of log(SFR$_{79}$) in the GV region may well be relatively narrow, with a dispersion of ≲0.2 dex, suggesting that any differences in the distribution of log(SFR$_{79}$) of the subdominant population of quenching objects and the dominant population of SFMS objects in the GV may be small.

We clearly need to try to subtract the contamination by SFMS galaxies before applying any number argument to derive $\tau_Q$. In principle, we could simply subtract from the GV population at a given value of SFR$_9$ the number of SFMS galaxies with the same SFR$_9$ but much higher SFR$_{79}$, i.e., effectively reflect the SFMS population around the locus of SFR$_{79}$ = 0 to remove it from the GV. Unfortunately, as noted in the previous section, the effect of the substantial noise in SFR$_{79}$ makes such an estimate of this contamination very difficult at low SFR$_9$. The diagonal error ellipses will be scattering the peak of the SFMS down to lower SFR$_9$ and higher SFR$_{79}$, spuriously increasing the number of SFMS counterparts of the GV galaxies without affecting the number of GV galaxies.

We could try to get around this problem with the following argument. Since the noise is symmetric in log(SFR$_{79}$) and is very small in log(sSFR$_7$), it has no effect on the mean log (SFR$_{79}$) of any sSFR$_7$-selected sample, including the overall sample of SF galaxies defined earlier to have Δlog (SFR$_7$) > −0.7. As noted above in Section 3.1, this sample has a very symmetric, roughly Gaussian distribution of log (SFR$_{79}$), with both the mean and median log(SFR$_{79}$) being ∼0. The measured scatter is ∼0.3 dex while the typical 1σ uncertainty is only ∼0.2 dex. The fact that the median log (SFR$_{79}$) is ∼0 in our sample of SF galaxies suggests that the putative number of truly quenching galaxies in that sample is being compensated by the elimination of truly SF galaxies that are undergoing a strong fluctuation in their SFR and are currently measured to have Δlog(SFR$_7$) < −0.7. If we further assume that the number of SF galaxies with Δlog (SFR$_7$) < −0.9 is negligible, this argument suggests that the number of quenching galaxies present in the overall population of galaxies with log(sSFR$_7$) > −0.9 is likely to be comparable to the observed number of galaxies observed within the GV, i.e., with −0.9 < log(sSFR$_7$) < −0.7.

Applying again the number argument from the quenching rate, but now considering the full range of SFR$_7$ rather than the 0.2 dex range of the GV, then yields a quenching timescale around ∼500 Myr, i.e., quite consistent with the value that was independently inferred above from the direct examination of the median log(SFR$_{79}$) of galaxies in the GV (but with the caveat noted above). In effect, the factor of 4.5 discrepancy that was noted above in the expected numbers of quenching galaxies *within* the GV strip is being compensated by the 4.5 times larger range in log(sSFR$_7$) of the overall SF population compared to the GV strip, i.e., 0.9 dex relative to 0.2 dex.

The convergence of these timescale estimates is suggestive (and less striking if the LINER correction in the number of GV galaxies is taken into account). However, we can again only claim (self-)consistency of our data with this timescale while noting that it is broadly consistent with those from Wetzel et al. (2013) and Hahn et al. (2017), who find that the e-folding quenching timescale is 200–800 Myr for satellites and 500–1500 Myr for central galaxies with generally decreasing timescales with increasing $M_*$.

It is clear that the main difficulty remains that of reliably identifying a set of objects that can be considered to be quenching. We stress again that, based on our own analysis, location within the SFR$_7$-defined GV is not by itself a sufficient condition for a galaxy to be considered to be "quenching." Such a GV galaxy may well have a counterpart (with the same long-term SFR$_9$) with a much higher SFR$_7$ (and SFR$_{79}$) and together form part of the stable SFMS population.

## 5. Summary and Conclusions

Galaxies are separated into two populations on the color–magnitude (or SFR–$M_*$) diagram: SF galaxies and quenched galaxies. Galaxies in between these two populations are usually called GV galaxies, and they are generally assumed to be transitioning from the SF to the quenched population, i.e., they are suffering from ongoing quenching processes. Observationally, it is quite challenging to tell whether this is the case or not, because one can usually only measure the current SFR of a given GV galaxy, rather than any tendency in its SFR.

In this work, we search for direct evidence of ongoing quenching processes in the galaxy population based on the star formation change parameter introduced by WL20. The Hα emission line traces the recent SFR within the last 5 Myr, and the Hδ absorption feature roughly traces the SFR within the last 800 Myr. Therefore, WL20 calibrated the star formation change parameter SFR$_{79}$, the ratio of the SFR on two different timescales SFR$_{5\,\mathrm{Myr}}$/SFR$_{800\mathrm{Myr}}$, based on these two spectral features plus an additional feature, the 4000 Å break. By definition, the SFR$_{79}$ directly tells us whether a given galaxy currently has an enhanced or a suppressed (s)SFR with respect to its (s)SFR averaged over the last ∼1 Gyr and therefore in principle provides a way to examine possible signatures of quenching in the galaxy population as well as quenching timescales.

Compared to the method in WL20, we make several improvements in calibrating the SFR$_{79}$ in the present work (see details in Section 2). The uncertainty of SFR$_{79}$ mainly comes from the uncertainty in measuring the absorption index of Hδ, due to the difficulty of decomposing Hδ emission and absorption. First, we therefore develop a new method to estimate the Hδ emission-line flux via the line fluxes of Hα and Hβ and show that this yields unbiased measurements of EW (Hδ$_A$) even for spectra of very low S/N (see Figure 2). Second, in calibrating the SFR$_{79}$, we use more realistic SFHs as input, including the possibility of quenching in order to account for potential quenching galaxies in the sample. Third, instead of using an analytic formula to calibrate the SFR$_{79}$ as in WL20, in this work, we construct and use a three-dimensional lookup table based on all the mock spectra. Fourth, we carefully assign the uncertainty in SFR$_{79}$ for each individual galaxy, including the observational uncertainty in the spectral features and the uncertainty in the calibration.

By applying the method described above to the 3″ fiber spectra from SDSS (Abazajian et al. 2009), we obtain the SFR$_{79}$ for each individual galaxy. Using this large galaxy sample, we first confirm the basic results found in WL20 for the SF population. First, the stability of the SFMS (the dispersion does not evolve with time) requires that the SFR$_{79}$ and the position of galaxies on the SFR$_9$-based SFMS are not correlated, which is indeed seen also in the present work (see the right panel of Figures 6 and 7). Second, we calculate the dispersion of the SFR$_7$-based and the SFR$_9$-based SFMS, as well as of the log (SFR$_{79}$) of SF objects. These dispersions contain information about the variability in the SFR of SF galaxies (Wang & Lilly 2020b). The dispersions as well as the qualitative trend of increasing dispersions with the stellar-mass surface density $\Sigma_*$ that we find in this work are overall very consistent with the results in WL20 despite the substantially larger and different data set used here. We then turn to look at objects significantly below the SFMS, in the traditionally defined GV, and search for direct evidence of quenching. We establish several new results in the present work:

1. The calibration of $SFR_{79}$ for objects below the SFMS is limited by several factors. The noise in the measurement of $EW(H\alpha)$ constitutes a fundamental limitation. In addition, the possible contribution to the $H\alpha$ emission from LINERs, which also becomes relatively more important for objects with low $H\alpha$ emission, further complicates the calibration of $SFR_{79}$ at low $EW(H\alpha)$. Finally, the calibration is affected by the choice of the prior in the distribution of SFHs used in the construction of the calibrator, not least the inclusion of quenching SFHs. In summary, the calibration of $SFR_{79}$ becomes more uncertain with decreasing $EW(H\alpha)$. Carefully examining all of the mentioned effects, we argue that the calibration should be reliable for $EW(H\alpha) > 4$ Å and therefore limit our subsequent analysis to galaxies above this limit.

2. We introduce the key diagram of $\log(SFR_{79})$ versus $\Delta\log(sSFR_9)$ for analysis of the data, in particular, to study any potential quenching signature (see Figure 7). When moving down in $sSFR_9$ to galaxies that are below the SF population (and galaxies in the GV) on this diagram, we clearly see asymmetries in the $\log(SFR_{79})$ distribution toward negative values. This is at least consistent with the presence of a genuine quenching population.

3. On this $\log(SFR_{79})$–$\Delta\log(sSFR_9)$ diagram, we show the tracks of model galaxies that are exponentially decreasing their SFRs with different characteristic e-folding timescales, assuming for simplicity that they start from the midpoint of the SF population. The observed galaxies below the SFMS are well covered by these quenching tracks, consistent with the idea that some fraction of these are quenching (see Figure 7).

4. Starting from the general assumption that the SFMS population should be stationary, i.e., that the distribution of the $sSFR_9$ of SFMS galaxies should not change with time, we derive the stationarity criterion for the distribution of $SFR_{79}$ for such a population to be mean$(SFR_{79}) - 1 = 0$ at any given $sSFR_9$. Since the noise in the data is symmetric in logarithmic space and will therefore bias the mean linear $SFR_{79}$ toward higher values, we instead approximate the stationarity criterion to be mean$(\log(SFR_{79})) = 0$ for our sample.

5. If there are genuinely quenching galaxies in the population, then we would expect to see deviations of the mean$(\log(SFR_{79}))$ toward negative values, as we move off the SFMS toward lower values of $\log(sSFR_9)$. We use the quenching formalism introduced in Peng et al. (2010) and Peng et al. (2012) to estimate quantitatively the size of this effect on the mean$(\log(SFR_{79}))$ as a function of $\Delta\log(sSFR_9)$. This depends somewhat on the average quenching timescales $\tau_Q$ and on the mass of the galaxies since different masses will have different overall quenching rates.

6. This prediction is however subject to two substantial observational effects that combine to reverse the predicted trend. First, the selection of objects with $EW(H\alpha) > 4$ Å introduces a strong bias toward higher $\log(SFR_{79})$ at low $\log(sSFR_9)$. Second, the significant noise in the $SFR_{79}$ measurements scatters objects along lines of constant $\log(sSFR_7)$, as can be seen from the error ellipses on the $\log(SFR_{79})$–$\Delta\log(sSFR_9)$ diagram (see Figure 8). The scattering of objects in the peak of the SFMS population therefore artificially increases the number of objects with low $\log(sSFR_9)$ but high (i.e., positive) $\log(SFR_{79})$ and thus counters the potential signature of quenching. These two effects combine to produce a predicted observational signal in reverse of what is expected for ongoing quenching, i.e., the mean$(\log(SFR_{79}))$ actually increases as $\Delta\log(sSFR_9)$ decreases. Differential effects with the quenching rate or quenching timescale should however be largely preserved, but are not easily distinguishable.

7. The observed mean $(\log(SFR_{79}))$ of galaxies is quite consistent with this modified prediction. We conclude that the distribution of galaxies on the $\log(SFR_{79})$–$\Delta\log(sSFR_9)$ diagram is certainly quite consistent with the presence of galaxies currently undergoing quenching but unfortunately cannot be used to establish this unequivocally.

8. If we naively assume that the galaxies below the SFMS, i.e., in the $SFR_7$-defined GV with $-0.7 > \Delta\log(sSFR_7) > -0.9$, are representative of the quenching population, then their median $\log(SFR_{79})$ (relative to the typical SFMS value) gives a direct estimate of a quenching timescale of $\tau_Q \sim 500$ Myr, largely independent of contamination from LINER emission.

9. However, the number of galaxies lying in the GV is a factor of 3–4.5 higher than expected from the standard Peng et al. (2010) and Peng et al. (2012) quenching rates for this same $\tau_Q$. This discrepancy suggests that the GV population is still dominated by galaxies that should better be considered to be the tail of the stable SFMS, as also indicated by inspection of the $\log(SFR_{79})$–$\Delta\log(sSFR_9)$ plot. These SFMS galaxies lying in the GV are the counterparts of indisputable SFMS galaxies that have the same $SFR_9$ but much higher $SFR_7$ (and thus $SFR_{79}$). Their number likely exceeds those of true "one-way" quenching galaxies, and we therefore caution against the presumption that the GV consists predominantly of "quenching" galaxies. This introduces a significant caveat to the quenching timescale derived from direct examination of the typical $SFR_{79}$ values of GV galaxies.

10. We can however try to estimate the total number of quenching objects in the overall SF population (integrated above the GV) and argue that this should in fact be quite similar to the total number of "SFMS plus quenching" galaxies that lie within the GV. It is essentially a coincidence that the factor of 3–4.5 (depending on the LINER correction) discrepancy within the GV in the previous bullet is being compensated by the 4.5× increase in the range of $\log(sSFR_7)$ if we consider the entire SF +GV instead of just the GV population. Reapplying the number argument, we now again get typical quenching timescales of ∼500 Myr, consistent with the previous direct estimates from the values of $SFR_{79}$. Again, we conclude that we can only claim (self-)consistency within the data for quenching timescales of this order.

We have tested that these conclusions are all robust to a maximum plausible LINER contribution to the $H\alpha$ emission (Section 2.3.4 and Appendix C). Neither are they affected by the ad hoc correction applied in Section 2.3.2 since we are concerned only about the values of log(SFR$_{79}$) relative to those of typical SFMS galaxies.

The star formation change parameter, SFR$_{79}$, is a powerful tool to study both the variability in the SFR of SF galaxies and quenching processes. Compared to almost all previous studies, it provides a different and valuable perspective on galaxy evolution on gigayear timescales in "real time." Despite the theoretical strength of this framework, it is disappointing that we nevertheless do not find unambiguous evidence of ongoing quenching processes in the galaxy population. This is partly due to the current limitations of the observational methodology, but also because, at least at the current cosmic epoch, the perturbation of the properties of the overall population of galaxies, by these "currently" quenching ones, is small. Even in the conventional SFR$_7$-defined GV, they are hard to disentangle from the much larger population of SFMS galaxies that may be undergoing strong but short-term fluctuations in their sSFR.

Nevertheless, if we assume that quenching is an ongoing physical phenomenon in the local universe and use simple estimates of the expected rate, then the distributions of SFR$_{79}$ that we find and analyze in this work consistently yield rather short e-folding quenching timescales of order 500 Myr.

## Acknowledgments

We thank the anonymous referee for their careful reading of the manuscript and for their constructive suggestions that improved the presentation of the paper.

This work was supported in part by the Swiss National Science Foundation (SNSF) through project grant 200020_207349. Funding for the SDSS and SDSS-II has been provided by the Alfred P. Sloan Foundation, the Participating Institutions, the National Science Foundation, the US Department of Energy, the National Aeronautics and Space Administration, the Japanese Monbukagakusho, the Max Planck Society, and the Higher Education Funding Council for England. The SDSS Web Site is http://www.sdss.org/. The SDSS is managed by the Astrophysical Research Consortium for the Participating Institutions. The Participating Institutions are the American Museum of Natural History, Astrophysical Institute Potsdam, University of Basel, University of Cambridge, Case Western Reserve University, University of Chicago, Drexel University, Fermilab, the Institute for Advanced Study, the Japan Participation Group, Johns Hopkins University, the Joint Institute for Nuclear Astrophysics, the Kavli Institute for Particle Astrophysics and Cosmology, the Korean Scientist Group, the Chinese Academy of Sciences (LAMOST), Los Alamos National Laboratory, the Max-Planck-Institute for Astronomy (MPIA), the Max-Planck-Institute for Astrophysics (MPA), New Mexico State University, Ohio State University, University of Pittsburgh, University of Portsmouth, Princeton University, the United States Naval Observatory, and the University of Washington.

## Appendix A
## The Variability in SFR on the SFMS

Wang & Lilly (2020b) developed a method to constrain the power spectrum distribution (PSD) of the time-variations in the SFR of galaxies on the SFMS, based on the dispersions of the SFMS$_7$ and the SFMS$_9$ ($\sigma_7$ and $\sigma_9$), as well as the dispersion in log(SFR$_{79}$) ($\sigma_{79}$). They found that $\sigma_{79}$ is closely related to the overall amplitude of the variations (i.e., the normalization of the PSD) while the ratio $\sigma_7/\sigma_9$ indicates the relative contribution of shorter and longer timescale variations to the overall dispersion of the SFMS (i.e., the slope of the PSD).

Here, we can investigate the dispersion in $\Delta$log(sSFR$_7$), $\Delta$log (sSFR$_9$), and log(SFR$_{79}$) with a much larger galaxy sample compared to that in Wang & Lilly (2020b). We separate the SF galaxies (defined as $\Delta$log(sSFR$_7$) > $-0.7$ dex; see Section 3.1) into four bins of total stellar mass according to Figure 1 as well as in four bins of stellar-mass surface density in the fiber, $\Sigma_*$. The latter are defined as $7.5 < \log(\Sigma_*/M_\odot \mathrm{kpc}^{-2}) < 8$, $8 < \log(\Sigma_*/M_\odot \mathrm{kpc}^{-2}) < 8.5$, $8.5 < \log(\Sigma_*/M_\odot \mathrm{kpc}^{-2}) < 9$, and $9 < \log(\Sigma_*/M_\odot \mathrm{kpc}^{-2}) < 9.5$. We show the three measured $\sigma$'s and the ratio $\sigma_7/\sigma_9$ as a function of stellar mass on the left and as a function of log($\Sigma_*$) on the right of Figure 11. We note that these dispersions are completely unaffected by the ad hoc correction (Section 2.3.2).

The shaded regions in Figure 11 correspond to the upper and lower limits of the intrinsic dispersion of the galaxy population. The lower limits are obtained by subtracting the typical uncertainties of the corresponding quantities in quadrature. For $\sigma_9$ and $\sigma_{79}$, we use the maximum estimate of the uncertainty in SFR$_{79}$, $u_{\mathrm{SFR}_{79},\mathrm{tot}}$, as discussed in Section 2.3.3. When inferring the uncertainty in sSFR$_9$, we do take into account that the uncertainties in SFR$_{79}$ and sSFR$_7$ are correlated (see Figure 8, below). The upper limits on the intrinsic dispersions are obtained by ignoring any uncertainty, i.e., they are the raw measured values from the data.

For comparison, we show the values of these three dispersions taken from WL20 as horizontal dashed lines in the left panel of Figure 11. We note that their quantities (i.e., SFR$_7$, SFR$_9$, and SFR$_{79}$) were measured within the effective radius of each galaxy. Further, the dispersions were computed for their entire SF sample without any binning and ignoring any noise contribution (which was however probably negligible in their case).

Overall, the dispersions found here, with a much larger and completely independent sample, are very consistent with those previously found in WL20, particularly if we account for noise (i.e., use the lower limits). Values of the $\sigma_7/\sigma_9$ ratio between 1 and 1.6 indicate a significant contribution of short-timescale SFR variations to the overall dispersion of the SFMS. While all the $\sigma$'s (and thus $\sigma_7/\sigma_9$) show almost no trend with $M_*$, they do slightly increase with $\Sigma_*$ in the higher two bins, again consistent with findings in WL20 where this correlation was interpreted as the response of a gas-regulator system to a time-varying inflow of gas. Assuming that $\Sigma_*$ traces the star formation efficiency (SFE; following, e.g., Shi et al. 2011), a higher $\Sigma_*$ will lead to a faster response to a given variation in the inflow and to more dispersion in the measured values of SFR$_{79}$. The trend found here is weaker than the trend found in WL20. This may follow from the fact that they were looking at a spatially resolved sample, taken from MaNGA (Bundy et al. 2015), enabling them to investigate annuli of individual galaxies while we are bound to the 3″ fibers of the SDSS, which always sample the inner regions of galaxies. The fraction of a given galaxy that is covered by the fiber then depends both on its physical size and its distance from the observer. Therefore, we are dealing with a smaller overall range of $\Sigma_*$ as compared to WL20, and it is hard to disentangle the different effects that determine the $\Sigma_*$ of a given object. For these reasons, we do not further investigate this trend and only point out the qualitative consistency between the results.

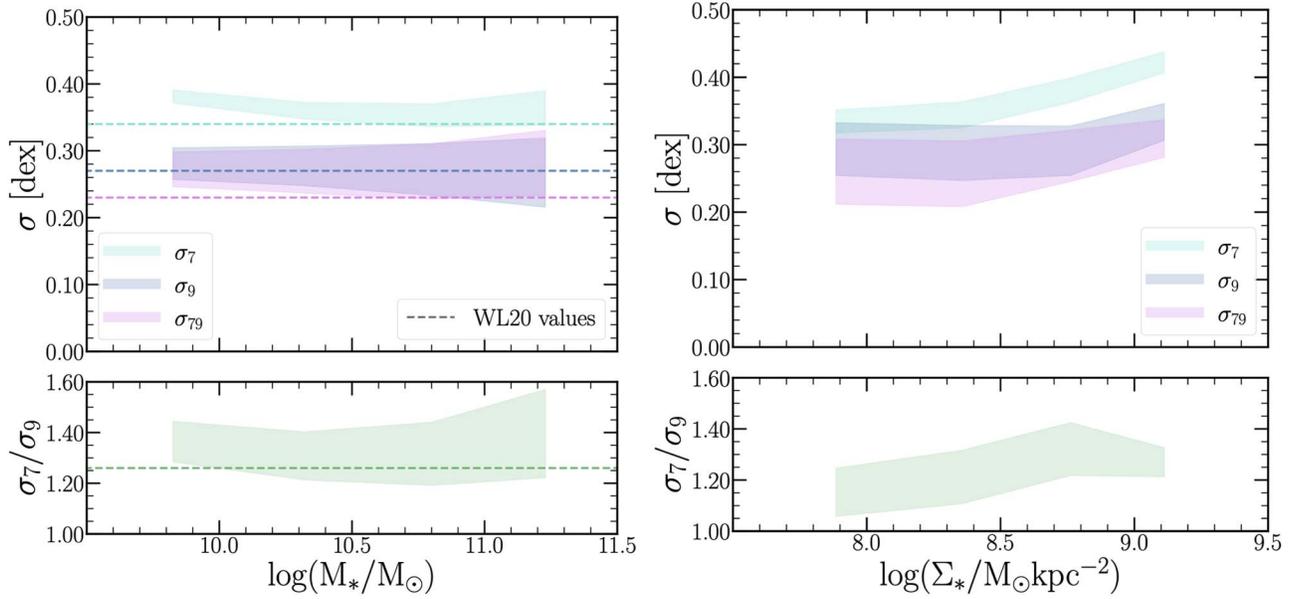

**Figure 11.** Measured dispersions of the SF population $\sigma_7$, $\sigma_9$, and $\sigma_{79}$ in the upper panel and the ratio $\sigma_7/\sigma_9$ in the lower panel as a function of stellar mass on the left and as a function of $\Sigma_*$ on the right, i.e., measured in four equally sized bins with a width of 0.5 dex. The shaded regions indicate the upper and lower limits of the dispersions (and their ratio), given by the raw measured value and the measured value with our maximum noise estimate subtracted in quadrature. The horizontal dashed lines in the left panel represent the values found in WL20 for the total dispersions of the sample of SF objects investigated there (without binning in mass).

The lack of correlation between $\Delta\log(\mathrm{sSFR}_9)$ and $\log(\mathrm{SFR}_{79})$ (see Section 3.2) further suggests that the dispersion of the $\mathrm{SFMS}_7$ is a combination of the dispersion present on longer timescales (i.e., $\sigma_9$) plus the additional effect of (uncorrelated) short-timescale fluctuations, as characterized by $\sigma_{79}$. We might therefore expect $\sigma_7^2 \approx \sigma_9^2 + \sigma_{79}^2$, and indeed this relation holds to within about 5% in all the $M_*$ and $\Sigma_*$ bins if we adopt our maximum noise estimates (i.e., assume the lower limits of the intrinsic dispersion respectively).

We have also examined the SFR time-variability for centrals and satellites separately. We find that overall the differences between the two populations are very small. In the lower two stellar-mass bins (i.e., for $M_* < 10^{10.5} M_\odot$), satellite galaxies appear to have slightly larger $\sigma_7$, $\sigma_9$, and $\sigma_{79}$ than centrals.

In addition, the dependence of the $\sigma$'s on $\Sigma_*$ appears to be weaker for satellites than for centrals. This may be interpreted as being due to satellite-specific processes, perhaps associated with satellite quenching. We will not discuss this in more detail, since it is not the main focus of this work.

## Appendix B
## Calculation of an Analytic Relation between $\mathrm{SFR}_{79}$ and $\tau_Q$

Based on the quenching model introduced in Section 2.3.1 and specifically Equation (1), we can compute the $\log(\mathrm{SFR}_{79})$ and $\Delta\log(\mathrm{sSFR}_7)$ for a given combination of $\tau_S$ and $\tau_Q$ as follows. Note that the same derivation is also valid for $\Delta\log(\mathrm{SFR}_{79})$, i.e., the $\log(\mathrm{SFR}_{79})$ relative to a (nonzero) reference value of the SF population. We can express the SFH of a quenching object as

$$\mathrm{sSFR}(\tau) = \mathrm{sSFR}_0 \begin{cases} 0 & \tau < \tau_S \\ \exp\left(\dfrac{\tau_S - \tau}{\tau_Q}\right) & \tau > \tau_S \end{cases} \quad (\mathrm{B1})$$

where $\mathrm{sSFR}_0$ is the nominal sSFR of the SFMS, i.e., $\Delta\log(\mathrm{sSFR}_7) = \log(\mathrm{sSFR}_7) - \log(\mathrm{sSFR}_0)$. Assuming that $\tau_S < \tau_0 - 5$ Myr, where $\tau_0 = 13{,}700$ Myr is the age of the universe, i.e., assuming that quenching started more than 5 Myr ago (which is a minor constraint given the quenching timescales considered), we can express $\mathrm{sSFR}_7$ as the average sSFR over the past 5 Myr as

$$\mathrm{sSFR}_7 = \frac{1}{5 \mathrm{\ Myr}} \int_{\tau_0 - 5\mathrm{\ Myr}}^{\tau_0} \mathrm{sSFR}(\tau) d\tau$$

$$= \frac{1}{5 \mathrm{\ Myr}} \int_{\tau_0 - 5\mathrm{\ Myr}}^{\tau_0} \mathrm{sSFR}_0 \exp\left(\frac{\tau_S - \tau}{\tau_Q}\right) d\tau$$

$$= ... = \frac{\mathrm{sSFR}_0\, \tau_Q}{5 \mathrm{\ Myr}} \exp\left(\frac{\tau_S - \tau_0}{\tau_Q}\right)\left(\exp\left(\frac{5 \mathrm{\ Myr}}{\tau_Q}\right) - 1\right). \quad (\mathrm{B2})$$

In order to derive an analogous expression for $\mathrm{sSFR}_9$, we have to distinguish two cases. In the case where quenching starts more than 800 Myr ago, i.e., $\tau_S < \tau_0 - 800$ Myr, a calculation completely analogous to Equation (B2) yields

$$\mathrm{sSFR}_9 = \frac{\mathrm{sSFR}_0\, \tau_Q}{800 \mathrm{\ Myr}} \exp\left(\frac{\tau_S - \tau_0}{\tau_Q}\right)\left(\exp\left(\frac{800 \mathrm{\ Myr}}{\tau_Q}\right) - 1\right). \quad (\mathrm{B3})$$

If however $\tau_S > \tau_0 - 800$ Myr, i.e., quenching started within the last 800 Myr, we need to compute $\mathrm{sSFR}_9$ as

$$\mathrm{sSFR}_9 = \frac{1}{800 \mathrm{\ Myr}}\left(\int_{\tau_0 - 800\mathrm{\ Myr}}^{\tau_S} \mathrm{sSFR}_0\, d\tau\right.$$
$$\left. + \int_{\tau_S}^{\tau_0} \mathrm{sSFR}_0 \exp\left(\frac{\tau_S - \tau}{\tau_Q}\right) d\tau\right)$$
$$= ... = \frac{\mathrm{sSFR}_0}{800 \mathrm{\ Myr}}\left(\tau_S - \tau_0 + 800 \mathrm{\ Myr} \right.$$
$$\left. + \tau_Q\left(1 - \exp\left(\frac{\tau_S - \tau_0}{\tau_Q}\right)\right)\right). \quad (\mathrm{B4})$$

Combining Equations (B2) to (B4), we obtain

$$\log(\mathrm{SFR}_{79}) = \log(\mathrm{sSFR}_7/\mathrm{sSFR}_9) = \log(160)$$

$$+ \begin{cases} \log\left(\dfrac{\exp\left(\frac{5\,\mathrm{Myr}}{\tau_Q}\right) - 1}{\exp\left(\frac{800\,\mathrm{Myr}}{\tau_Q}\right) - 1}\right) & \tau_S < \tau_0 - 800\,\mathrm{Myr} \\ \log\left(\dfrac{\exp\left(\frac{5\,\mathrm{Myr}}{\tau_Q}\right) - 1}{\left(\frac{\tau_S - \tau_0 + 800\,\mathrm{Myr}}{\tau_Q} + 1\right)\exp\left(\frac{\tau_0 - \tau_S}{\tau_Q}\right) - 1}\right) & \tau_S > \tau_0 - 800\,\mathrm{Myr} \end{cases}$$

(B5)

Since we did not find an analytic inverse of this function, we use a numeric approximation to get from a given combination of $\Delta\log(\mathrm{sSFR}_7)$ and $\log(\mathrm{SFR}_{79})$ to a $(\tau_S, \tau_Q)$. We start with a range of values of $\tau_Q$, for each of which we compute a corresponding $\tau_S$ at a given $\Delta\log(\mathrm{sSFR}_7)$ and using Equation (B2), which can be solved for $\tau_S$. For each tuple $(\tau_S, \tau_Q)$, we can then compute $\log(\mathrm{SFR}_{79})$ using Equation (B5). In this way, we can construct a lookup table, matching tuples $(\tau_S, \tau_Q)$ to values of $\log(\mathrm{SFR}_{79})$ at fixed $\Delta\log(\mathrm{sSFR}_7)$. From the lookup table corresponding to the measured $\Delta\log(\mathrm{sSFR}_7)$, we finally find the $(\tau_S, \tau_Q)$ that best matches the measured $\log(\mathrm{SFR}_{79})$ of a given galaxy or galaxy population.

We show the conversion functions $(\Delta)\log(\mathrm{SFR}_{79}) \to \log(\tau_Q/\mathrm{Gyr})$ assuming $\Delta\log(\mathrm{sSFR}_7) = -1.2, -1, -0.8, -0.6, -0.4, -0.2$ dex in different colors and line styles in Figure 12.

So far, we have assumed that quenching starts from the midpoint of the SF population at $\Delta\log(\mathrm{sSFR}_7) = \log(\mathrm{SFR}_{79}) = 0$. However, this may not be true in the real universe where objects might have their sSFR suppressed for some time before the actual quenching process starts, or they may experience a phase of enhanced star formation (a starburst) prior to quenching. Any such scenario would likely produce a more complicated quenching track on the $\log(\mathrm{SFR}_{79})$–$\Delta\log(\mathrm{sSFR}_9)$ diagram but to give a rough idea of how the quenching timescales derived from the typical value of $\log(\mathrm{SFR}_{79})$ of GV galaxies are affected by a change of the starting point of the quenching tracks, we note that shifting the starting point of the tracks up (down) is equivalent to measuring $\tau_Q$ at a lower (higher) value of $\Delta\log(\mathrm{sSFR}_7)$. For example, for an object that has a measured $\Delta\log(\mathrm{sSFR}_7)$ of $-0.8$, the conversion function $(\Delta)\log(\mathrm{SFR}_{79}) \to \tau_Q$ shown as the orange dotted line in Figure 12 is the one we would use if quenching starts from the midpoint of the SFMS. If however, quenching started 0.4 dex above (below) the SFMS, then the accordant conversion function would be the brown dashed–dotted (blue dashed) line corresponding to $\Delta\log(\mathrm{sSFR}_7) = -1.2$ ($\Delta\log(\mathrm{sSFR}_7) = -0.4$). Note that by construction, the minimum $\Delta\log(\mathrm{SFR}_{79})$ that can be measured at a fixed $\Delta\log(\mathrm{sSFR}_7)$ is equal to the latter for an object quenching along a quenching track. So, any direct comparison between inferred timescales in Figure 12 is only meaningful vertically, i.e., at a fixed $\Delta\log(\mathrm{SFR}_{79})$ and between the conversion functions that actually cover that value. This illustrates that for the $\Delta\log(\mathrm{SFR}_{79}) = -0.38$ that we find for the GV galaxies in our sample, choosing the starting point 0.4 dex above the SFMS midpoint does not affect the inferred median quenching timescale of $\sim 500$ Myr, but it would lengthen inferred timescales if we had measured a lower $\Delta\log(\mathrm{SFR}_{79})$ of e.g., $-0.7$. On the other hand, the conversion function corresponding to the starting point below the SFMS would yield a median quenching timescale between 100 and 300 Myr in our sample, i.e., it would lead to shorter timescales. We emphasize again that this is not a realistic scenario but only serves to illustrate the potential effect of a different starting point of quenching on the inferred quenching timescales.

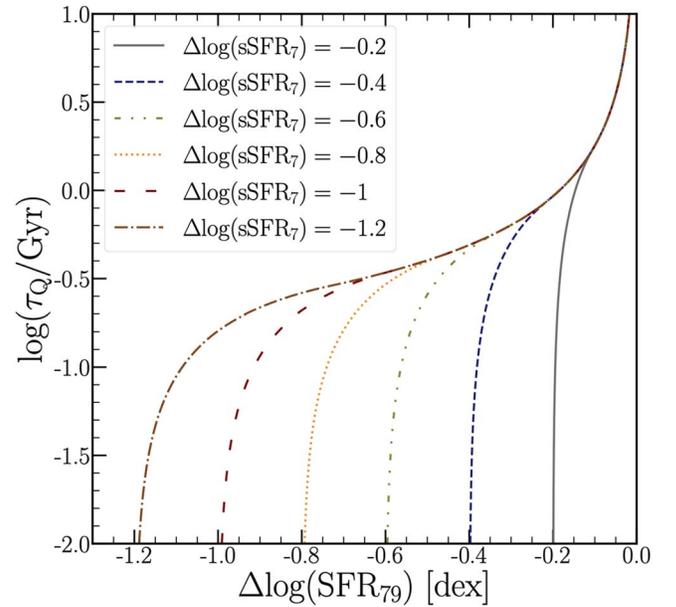

**Figure 12.** Conversion function used to convert $\Delta\log(\mathrm{SFR}_{79})$ into a quenching timescale $\tau_Q$ based on the analytic relation in Equation (B5). Different colors and line styles represent different values of $\Delta\log(\mathrm{sSFR}_7)$ at which the conversion is performed or, equivalently, different starting points of the quenching tracks above or below the SFMS as explained in the text.

### Appendix C
### Constraining the Effect of LINER Emission on Our Results

In order to investigate the effect of LINER emission on our results, we correct our measured and dust-corrected H$\alpha$ emission, assuming a maximum plausible contribution of LINER emission. The idea behind this correction is therefore not to realistically correct for LINER emission but to illustrate the maximum effect that LINERs might have on our results.

On the left panel of Figure 13, we show the $\log(\mathrm{N\,II}/\mathrm{H}\alpha)$–$\log(\mathrm{O\,III}/\mathrm{H}\beta)$ distribution of our sample. We adopt the thresholds given in Kewley et al. (2006) to distinguish between SF objects, composites, and AGNs. The colored markers on the plot show the median line ratios in our four mass bins and in bins of $\Delta\log(\mathrm{sSFR}_7)$ indicated by the color coding. This already makes the point that with decreasing $\Delta\log(\mathrm{sSFR}_7)$ and more distinctly at higher $M_*$, the typical object in our sample moves from the SF through the composite- and close to the AGN region of the diagnostic diagram. Another illustration of the effect is shown on the right panel of Figure 13, where the fraction of objects belonging to each of the three groups (SF, composite, or AGN) is shown as a function of $\Delta\log(\mathrm{sSFR}_7)$. While the fraction of objects classified as SF monotonically decreases, the fraction of composites and AGNs both increase with decreasing $\Delta\log(\mathrm{sSFR}_7)$. This indicates that there might indeed be a significant contribution of LINERs to our measured H$\alpha$ emission at low $\Delta\log(\mathrm{sSFR}_7)$.

We now proceed as follows. Based on the work of Belfiore et al. (2016), we assume that the maximum contribution of LINERs to the dust-free EW(H$\alpha$) of an individual galaxy is 3 Å. For all composite objects, we then reduce their measured and dust-corrected EW(Hα) by a factor of 2, but not by more than 3 Å, and for all objects classified as AGNs, we directly reduce their EW(Hα) by the maximum LINER contribution of 3 Å. We emphasize that this is a somewhat arbitrary correction that serves to illustrate the possible effect of LINERs on our results. We add dust back into the corrected EW(Hα), inverting our dust correction prescription (Section 2.2), and multiply the dusty EW(Hα) with the measured continuum of each spectrum to retrieve a LINER-corrected, dusty Hα flux, which we then again correct for dust to get an updated estimate of $sSFR_7$ adopting the star formation law of Kennicutt & Robert (1998). We use the LINER-corrected estimate of the dust-free EW(Hα) together with the other spectral features, which we leave unaltered to derive a new $SFR_{79}$ using our calibrator. Figure 14 shows the comparison of the overall $\log(SFR_{79})$–$\Delta\log(sSFR_9)$ diagram without (left panel, a duplicate of Figure 7) and with (right panel) the LINER correction.

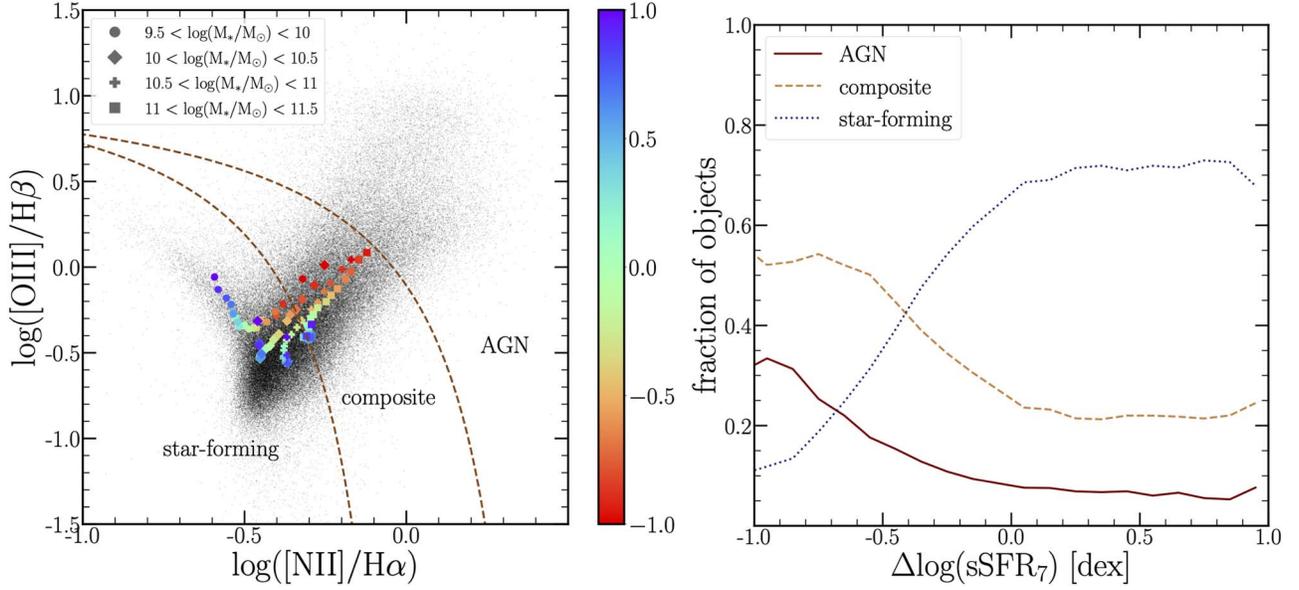

**Figure 13.** On the left, we show the log (N II/Hα)–log (O III/Hβ) distribution of our sample. We adopt the brown dashed lines from Kewley et al. (2006) to distinguish between SF objects, composites, and AGNs. The colored markers on the plot show the median line ratios in four mass bins and in bins of $\Delta\log(sSFR_7)$ indicated by the color coding. It is apparent from this median relation that with decreasing $\Delta\log(sSFR_7)$ and more distinctly at higher $M_*$, the typical object in our sample moves farther toward the AGN region on the diagram. On the right, we show the fraction of objects in our sample belonging to each of the three categories defined in the left panel as a function of $\Delta\log(sSFR_7)$ to further illustrate the significance of LINERs in the regime where we define our GV ($-0.9$ dex $< \Delta\log(sSFR_7) < -0.7$ dex).

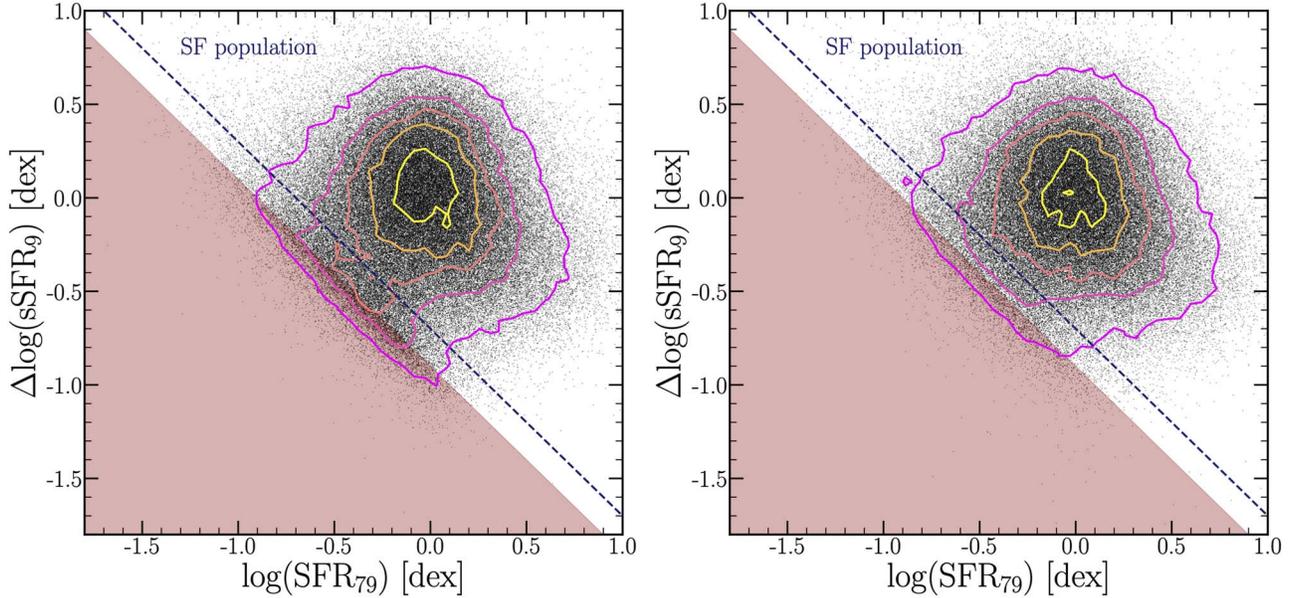

**Figure 14.** $\log(SFR_{79})$–$\Delta\log(sSFR_9)$ diagram for our full sample without the LINER correction on the left (a duplicate of Figure 7) and with the LINER correction on the right for comparison.

## Appendix D
### Effects of the Prior Distribution of SFHs

Our method of calibrating $SFR_{79}$ involves a prior in the sense that we start with a predefined range of SFHs, which we then use to produce spectra to extract the corresponding spectral features that will be matched to the observations. Our approach to this was to start with a very broad range of SFHs to cover all possible SFHs in the universe by superimposing stochastic fluctuations on the otherwise smooth SFHs that follow the cosmic evolution of the SFMS. While this may lead to an overestimate of the uncertainty in the derived $SFR_{79}$ intrinsic to the calibrator, as discussed in Section 2.3, we believe that it allows a relatively unbiased determination of the values of $SFR_{79}$. The situation is however different when it comes to quenching. In order to account for objects in the data that show relatively low H$\alpha$ emission and H$\delta$ absorption, it is inevitable to include SFHs in which the SFR has been suppressed for a sufficiently long time, i.e., it decreases on some timescale and then either keeps decreasing or remains at a suppressed level. As described in Section 2.3, we experimented with two implementations of the quenching process, which we superimpose on half of the SFHs. In the first implementation, the SFR declines exponentially "forever" (subsequently referred to as the quenching prior A), and in the second, we set a "floor" to the quenching process, i.e., the SFR of a quenching galaxy decreases by 1.3 dex and then remains at the corresponding low and constant value (subsequently referred to as the quenching prior B). We have also tested different values for the "floor," such as 0.9 dex or 2 dex below the SFMS, and found only minor differences between them. With the quenching prior A, sample objects with low H$\alpha$ emission and H$\delta$ absorption are predominantly matched with SFHs in which the SFR is still declining exponentially and therefore have a negative log($SFR_{79}$). In the calibration, we are selecting mock galaxies that constitute a three-dimensional Gaussian distribution in the spectral features around the measured values with dispersions given by the measurement uncertainties (see Section 2.2). As we approach lower and lower values of EW(H$\alpha$), this will tend to include SFHs all the way down to EW(H$\alpha$) = 0 Å and therefore the retrieved value of $SFR_{79}$ will depend on the distribution of quenching timescales $\tau_Q$ that we put in. Since we used a uniform distribution of $\tau_Q$ in logarithmic space, this is dominated by short timescales yielding very low values of $SFR_{79}$. With the quenching prior B, "quenched" mock galaxies are fluctuating around a constant SFR (due to the stochastic fluctuations that are still superimposed), similar to SF mock galaxies, but at a suppressed level of star formation. So, in principle, their typical log($SFR_{79}$) is expected to be ~0. In any model that involves stochastic fluctuations in the SFR, the lowest values of EW(H$\alpha$) will however always be associated with local minima in the mock SFHs and thus with negative log($SFR_{79}$). Therefore, sample objects with low EW(H$\alpha$) and EW(H$\delta_A$) still typically get a negative value of log($SFR_{79}$) assigned, which is however closer to 0 than with the quenching prior A. Note that by "low" H$\alpha$ emission or H$\delta$ absorption we mean something like EW(H$\alpha$) $\lesssim$ 4 Å, which we introduced as a boundary on the reliability of the calibration. Most of those objects in the data also have a low EW(H$\delta_A$) $\approx$ −2 Å with the bulk of the objects between −4 and 1 Å. There is a small number of objects with very low or even vanishing EW(H$\alpha$), which do however show significant H$\delta$ absorption, e.g., there are 2685 objects with EW(H$\alpha$) < 4 Å and EW(H$\delta_A$) > 2 Å, corresponding to ~1% of the entire sample. Those objects have likely quenched recently and quite rapidly and/or have quenched after a burst of star formation.

The bulk of the objects with EW(H$\alpha$) < 4 Å have likely quenched a relatively long time ago, and it is therefore no surprise that our calibration does not work for those objects. To illustrate the effects discussed in this section, we show the log($SFR_{79}$)–$\Delta$log($sSFR_9$) diagram for all of our sample galaxies in the two top panels in Figure 15. All four panels are produced in analogy to Figure 7. The magenta contours enclose 10%, 30%, 50%, 70%, and 90% of the sample galaxies. In the top-left panel of Figure 15, we show our results as obtained using the quenching prior A, and in the top right, we show the same objects but for the quenching prior B. Note that it is this latter version of the calibration that we eventually used throughout the paper. The two bottom panels then show analogous plots but only display objects with EW(H$\alpha$) > 4 Å. For a better illustration of that sample selection, we color the region corresponding to $\Delta$log($sSFR_7$) < −0.9 (roughly equivalent to EW(H$\alpha$) > 4 Å) in red. Figure 15 shows that the quenched objects in the sample form a diagonal sequence that extends from the bottom right to the upper left of the diagram and whose exact shape and location are extremely sensitive to the choice of a prior distribution of SFHs. For the quenching prior A, the quenched population is overall shifted up and to the left. Note that this is simply a consequence of this prior yielding lower values of $SFR_{79}$ for those objects as discussed above, which then also directly translate to higher values of $\Delta$log($sSFR_9$) since $\Delta$log($sSFR_7$) is obtained from the H$\alpha$ luminosity, independent of the $SFR_{79}$ calibration. Note also that 16% of the sample objects are still missing from that plot because they are classified as having EW(H$\alpha$) = 0 Å (see Section 2.2). The lower two panels of Figure 15 then demonstrate that for EW(H$\alpha$) > 4 Å, the calibration is fairly robust to the choice of a prior distribution of quenching SFHs, justifying the adoption of this cut.

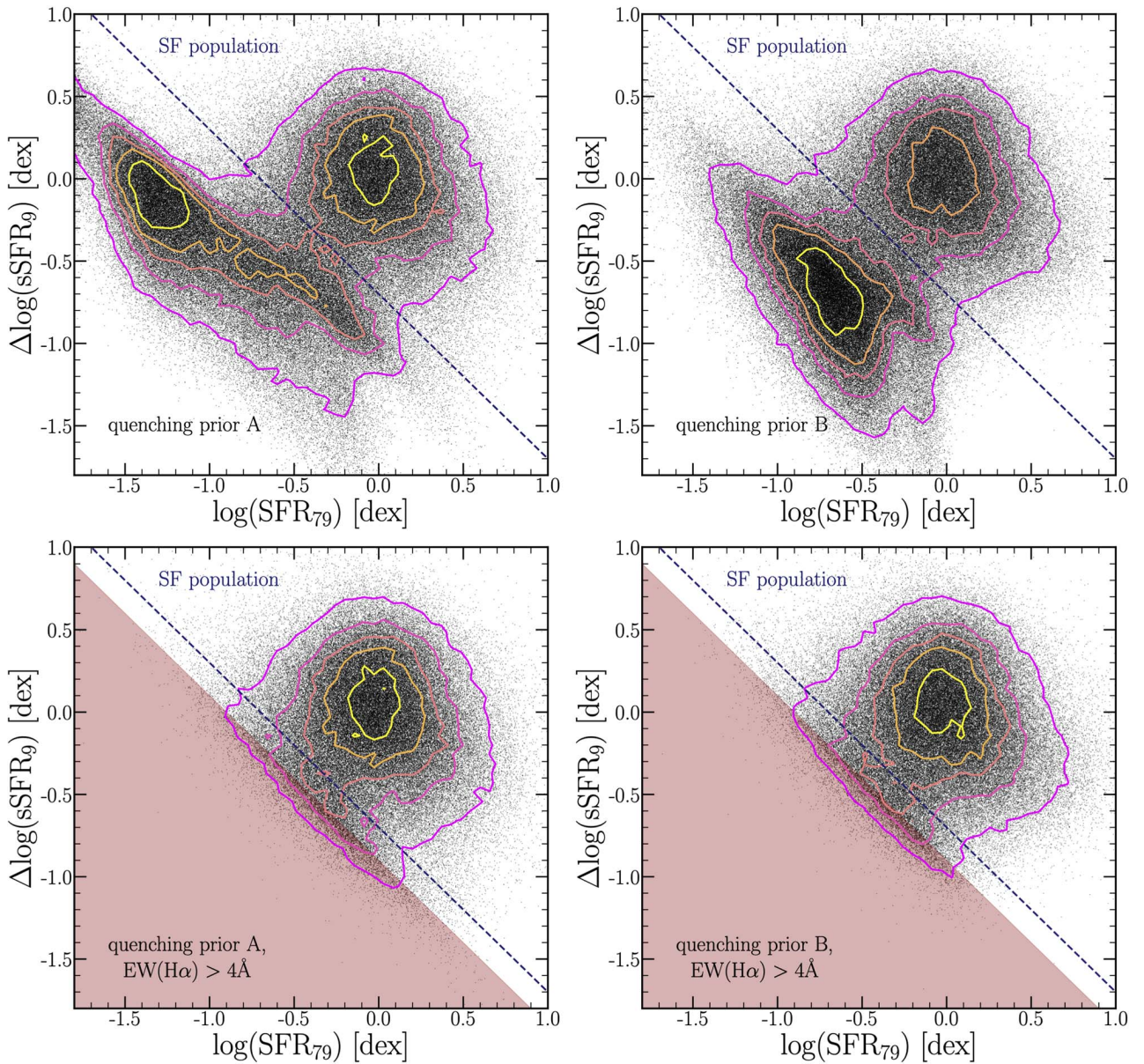

**Figure 15.** Four versions of the log(SFR$_{79}$)–Δlog(sSFR$_9$) diagram produced largely in analogy to Figure 7. The black dots represent a simple scatter plot of our entire initial sample in the top panels and of objects with EW(Hα) > 4 Å in the bottom panels. The colored solid lines indicate the contours including 10%, 30%, 50%, 70, and 90% of the underlying black dots. In the left panels, we show the resulting distributions if we assume a prior distribution of quenching SFHs for the calibration in which the SFR declines exponentially "forever" (quenching prior A), and in the right panels, we show the same distributions if we assume a prior where there is a "floor" to the quenching process 1.3 dex below the SFMS (quenching prior B; see the text for details). The blue dashed line represents the lower boundary of the SF population defined in Section 3.1, and the red shaded area in the bottom panels marks the region of the diagram where Δlog(sSFR$_7$) < −0.9, which is roughly equivalent to EW(H α) < 4 Å and makes the sample selection more clear.

## Appendix E
## Effects of an Added Old Stellar Population

In the following, we investigate the effect of adding a substantial OSP to a typical SF galaxy on its location on the log(SFR$_{79}$)–Δlog(sSFR$_9$) diagram. This effectively simulates galaxies with very different SFHs compared to those considered above in the context of both SF as well as quenching galaxies. By adding a substantial OSP to a continually SF galaxy, we simulate a galaxy that formed a large fraction of its stars in a burst some significant time ago but thereafter has maintained a more or less constant SFR at a substantially sub-SFMS level. Such a galaxy might lie below the SF population because of its low sSFR but should probably not be considered "quenching" as it does not have a currently declining SFR. Where would such an object lie in the log(SFR$_{79}$)–Δlog(sSFR$_9$) plane, and could it be misidentified to lie at the bottom left of the SF cloud where we argued to see some indication for ongoing quenching?

Using the FSPS code (Conroy et al. 2009), we model the spectral features of a single stellar population at different ages adopting the MILES stellar library (Sánchez-Blázquez et al. 2006; Falcón-Barroso et al. 2011), a Chabrier (2003) IMF, and the Padova isochrones (e.g., Bertelli et al. 1994, 2008) as in Section 2.3.1. We then start with the spectral features of a typical SF galaxy from our sample (i.e., a typical galaxy around Δlog(sSFR$_7$) = log(SFR$_{79}$) = 0) and successively add a contribution of the modeled OSP adopting different ages and and to any (sub)set of galaxies on the log(SFR$_{79}$)–Δlog(sSFR$_9$) diagram.

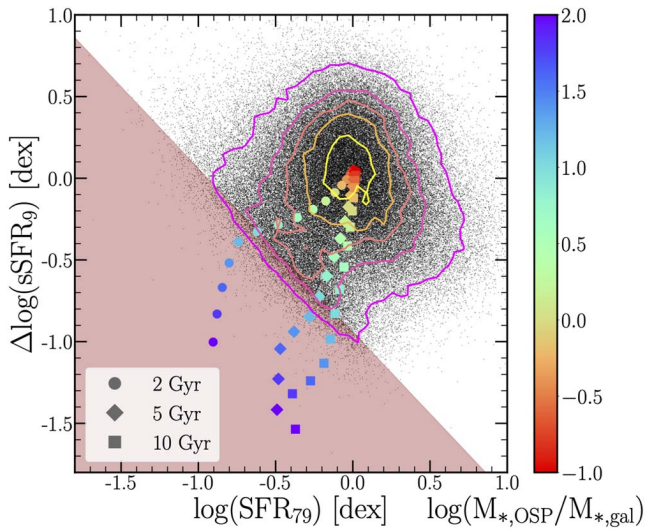

**Figure 16.** log(SFR$_{79}$)–Δlog(sSFR$_9$) diagram for the galaxy sample with EW(Hα) > 4 Å in analogy to Figure 7. The overplotted markers indicate the derived values of log(SFR$_{79}$) and Δlog(sSFR$_9$) for a typical SF galaxy (log(SFR$_{79}$) = Δlog(sSFR$_9$) = 0) with the added contribution of an OSP of different ages (2, 5, and 10 Gyr, illustrated by the marker types) and mass fractions (equally spaced values in log, ranging from −1 to 2, illustrated by the color coding).

## ORCID iDs

Andrea Weibel https://orcid.org/0000-0001-8928-4465
Enci Wang https://orcid.org/0000-0003-1588-9394
Simon J. Lilly https://orcid.org/0000-0002-6423-3597

mass fractions with respect to the initial mass of the SF population. We select the initial mass of the SF galaxy to be $10^{10.5}M_\odot$, roughly the typical mass of an SF galaxy in our sample. We then investigate how the estimates of the star formation parameters sSFR$_7$, SFR$_{79}$, and thus also sSFR$_9$ are affected by the increasing contribution of an OSP. In principle, we would expect sSFR$_7$ and sSFR$_9$ to scale exactly inversely with the added mass, while SFR$_{79}$ should remain unaltered. In other words, we would ideally see the composite galaxy move vertically downwards in the log(SFR$_{79}$)–Δlog(sSFR$_9$) diagram as more OSP is added.

The results of this exercise are shown in Figure 16, where 15 different mass fractions (equally spaced in log, ranging from −1 to 2, illustrated by the color coding) are plotted for three different ages of the OSP (2, 5, and 10 Gyr, illustrated by different marker types).

As might be expected, the older the OSP population is, the better the composite galaxy follows the desired vertical track in the log(SFR$_{79}$)–Δlog(sSFR$_9$) diagram.

For all ages, an OSP with a mass roughly 2–3 times the mass of the SF component is required to move an object out of the SF population. For the older OSP ages, 5 Gyr and greater, such an object would not mimic a galaxy with a currently significantly suppressed sSFR because it (correctly) still has log(SFR$_{79}$) ∼ 0. For younger ages of the added OSP, the composite galaxy may well contaminate the region to the bottom left of the SF population with log(SFR$_{79}$) ≈ −0.4.

However, a galaxy with an OSP that is 2–3 times as massive as the population of continually formed stars but with an age of only 2–5 Gyr would represent a rather odd SFH. It would have effectively required 65%–75% of the stellar mass to have been formed in a short burst a few gigayears ago. While such galaxies may exist, we would not expect them to be very common, and they are therefore unlikely to contribute significantly to the population of galaxies to the bottom left of the SF population